\documentclass[journal=jctcce,manuscript=article]{achemso}

\usepackage{chemformula} % Formula subscripts using \ch{}
\usepackage{hyperref}
\usepackage{graphicx}
\usepackage{dcolumn}
\usepackage{amsmath}
\usepackage{amssymb}
\usepackage{bigints}
\usepackage{array}
\usepackage{tabularx}
\usepackage{multirow}
\usepackage{booktabs}
\usepackage{siunitx}
\usepackage{float}
\usepackage[caption=false]{subfig}
\usepackage{bm}
\usepackage{cases}
\usepackage[utf8]{inputenc}
\usepackage[T1]{fontenc}
\usepackage{xcolor}
\usepackage{soul}
\usepackage{tikz}

\newcommand{\br}{\mathbf{r}}
\newcommand{\bp}{\mathbf{p}}

\newcommand{\bff}{\mathbf{f}}  
\newcommand{\be}{\mathbf{e}} 
 
\newcommand{\bx}{\mathbf{x}}

\newcommand{\occ}{\bar{c}}

\newcommand{\tT}{\tilde{\theta}}
\newcommand{\tP}{\tilde{\pi}}
\newcommand{\tS}{\tilde{s}}

\newcommand{\Rc}{R_\text{cut}}
\newcommand{\tRc}{\tilde{R}_\text{cut}}

\newcolumntype{Y}{>{\centering\arraybackslash}X}

\newenvironment{disclaimer}{
  \section*{Disclaimer}
}{}

\author{Giuseppe Colella}
\affiliation{Departament d'Enginyeria Química, ETSEQ, Universitat Rovira i Virgili, Tarragona, 43007 Spain}
\author{Allan D. Mackie}
\affiliation{Departament d'Enginyeria Química, ETSEQ, Universitat Rovira i Virgili, Tarragona, 43007 Spain}
\author{James P. Larentzos}
\affiliation{U.S. Army Combat Capabilities Development Command (DEVCOM) Army Research Laboratory, Aberdeen Proving Ground, MD, 21005 USA}
\author{Fernando Bresme}
\affiliation{Department of Chemistry, Molecular Sciences Research Hub, Imperial College London, London W12 0BZ, UK and Thomas Young Centre for Theory and Simulation of Materials,
Imperial College London, London SW7 2AZ, United Kingdom}
\author{Josep Bonet Avalos}
\email{josep.bonet@urv.cat}
\affiliation{Departament d'Enginyeria Química, ETSEQ, Universitat Rovira i Virgili, Tarragona, 43007 Spain}

\title{Statistical Mechanics of Density- and Temperature-Dependent Potentials: Application to Condensed Phases within GenDPDE}

\begin{document}

\begin{abstract}
    Coarse-grain Lagrangian methods, such as Dissipative Particle Dynamics (P. J. Hoogerbrugge {\em et al.}, {\em EPL}, \textbf{1992}, \textit{19}, 155), are suitable to describe mesoscopic fluid systems with the inclusion of thermal fluctuations. However, the realistic simulation of liquids using these methods represents a longstanding problem. In this work, we develop a local thermodynamic (LTh) model for the description of condensed phases within the framework of the Generalized Dissipative Particle Dynamics with Energy Conservation (GenDPDE) method (J. Bonet Avalos {\em et al.}, {\em Phys. Chem. Chem. Phys.} \textbf{2019}, \textit{21}, 24891). Such a model is appropriate for the analysis of liquids, due to the explicit account of the thermal expansion coefficient and isothermal compressibility at the mesoscale. We demonstrate the accuracy of the LTh model by inspecting the thermodynamic properties of argon at both liquid and supercritical conditions, through equilibrium simulations carried out around two characteristic reference states ($T=125.7$ K, $P=85.31$ MPa, $\tilde{\rho} = 1419.7$ kg/m$^3$ for liquid Ar, and $T=418.8$ K, $P=85.31$ MPa, $\tilde{\rho} = 695.99$ kg/m$^3$ for supercritical Ar). Remarkably, we show that the model is also valid in a range of thermodynamic conditions near the reference states, allowing for a correct description of the physics of systems with spatial density and temperature variations. We furthermore derive analytical expressions for the macroscopic pressure and energy equations of state in terms of the model parameters, discussing their validity and limitations. We show that, even at the mean-field level, a correct account of the local particle arrangements is necessary to obtain accurate predictions of the macroscopic thermodynamic quantities from mesoscopic properties. Thus, we also investigate the applicability of the Hypernetted Chain approximation as a tool to predict the radial distribution function of the GenDPDE system, examining the strengths and deficiencies of this approach. With the proposed LTh model, GenDPDE provides a reliable and flexible tool for the analysis of condensed phases through coarse-grain techniques.
\end{abstract}

\maketitle

\section{Introduction}

The use of coarse-grain (CG) numerical approaches to the study of the equilibrium and dynamic properties of complex systems, such as colloidal suspensions \cite{Padding2006}, polymeric solutions \cite{Locatelli2016}, and biomolecular systems \cite{Noid2013, Pak2018}, among others, has attracted great attention in recent years. These systems usually involve a large number of physical constituents and are characterized by time and length scales that are much larger than molecular dimensions. As a consequence, the application of atomistic techniques like Molecular Dynamics (MD) to their analysis is often computationally unfeasible. CG methods thus represent a valuable alternative to simulate cases in which atomistic approaches are impractical.

Among the available CG methods, Dissipative Particle Dynamics (DPD) \cite{Hoogerbrugge1992,Espanol1995,Groot1997} is a popular tool to deal with a variety of relevant problems \cite{Moeendarbary2009,Moeendarbary2010,Espanol2017, Santo2021}. However, despite its wide range of applicability, DPD also presents several limitations. First, as a dissipative model, it does not ensure energy conservation, and can thus only be applied to isothermal situations. Moreover, in its original formulation, DPD cannot reproduce liquid-vapor coexistence, due to the form of the associated Equation of State (EoS), which is quadratic in the density \cite{Groot1997}. Extensions of the method have thus been developed to model more complex EoS, via the definition of a potential energy that depends on many-body rather than pairwise interactions, through a {\em local density} \cite{Pago2000,Warren2003,Zhao2021}. These many-body DPD (MDPD) approaches, however, are still limited to isothermal scenarios and cannot describe heat flux.

The Generalized Dissipative Particle Dynamics with Energy conservation (GenDPDE) method \cite{Avalos2019, Avalos2021} has recently been introduced to overcome this latter constraint. GenDPDE is based on an {\em internal} thermodynamic description for the CG-ed particles (often referred to as {\em mesoparticles}), that characterizes its internal state as a function of the resolved degrees of freedom (DoF). Mesoparticles are thus seen as property carriers with internal energy, stored in the non-resolved degrees of freedom (DoF). This internal description for the mesoparticles makes it possible to consistently define density- and temperature-dependent potentials, which are relevant for the analysis of complex non-equilibrium phenomena, beyond the capacity of the traditional DPD model \cite{Lisal2022a,Lee2023}.

The inherent flexibility of the GenDPDE framework also permits the introduction of particle thermodynamic models suitable for the description of liquid phases, through the definition of the relevant parameters at the mesoscopic scale. The analysis of liquids through DPD-like CG models has actually proven to be challenging when the goal is to make quantitative predictions of the system properties (see, e.g., Ref. \cite{Sokhan2023}). Thus, the aim of the present work is, on the one hand, to introduce a simple local thermodynamic (LTh) model capable of describing condensed phases. On the other hand, we calibrate the LTh model parameters from macroscopic equations of state (EoS), using as input the thermophysical properties of a real fluid of interest.  We show that this procedure allows us to both qualitatively and quantitatively simulate condensed phases using the GenDPDE method.

%\FB{These articles may be of interest for the discussion above:\href{https://www.researchgate.net/publication/325301734_Dissipative_particle_dynamics_Dissipative_forces_from_atomistic_simulation} \href{https://pubs.rsc.org/en/content/articlehtml/2023/sm/d3sm00835e}. The second one goes very much in the direction of modelling liquid phases; in fact, they model coexisting curves. Also see the comparison with the IAPWS reference in Figure 7. They do water.}

In the derivation of the macroscopic EoS, we have extended the principles of Equilibrium Statistical Mechanics to density- and temperature-dependent potentials. To the best of our knowledge, this analysis has never been explored before. Although our approach embraces the classical theory of liquids \cite{Hansen2006}, and previous analogous studies on many-body potentials \cite{Merabia2007}, the complexity of the latter goes far beyond that of typical pairwise counterparts. Remarkably, our analysis reveals that the local mesoparticle arrangements play a crucial role in accurately determining the relationship between macroscopic and mesoscopic properties, i.e., in the reproducibility of macroscopic properties from a CG mesoscopic model. In a previous study (see Ref. \cite{Colella2025}), we found that mesoscopic systems described by density-dependent potentials are prone to form spurious local structures, particularly when simulating the liquid phase, where potential interactions dominate over kinetic effects. To address this issue, we introduced a redefinition of the mesoparticle volume, that minimizes the presence of such substructures and allows us to recover meaningful results for simple, well-behaved liquids such as argon. However, we also found that in more complex fluids, like e.g. water, the formation of spurious substructures is essentially unavoidable. For this reason, in the present work we choose to focus again on argon as our test substance. We thus demonstrate the validity and usefulness of the proposed LTh model by analyzing the behavior of the system both at the targeted reference state as well as in its close surroundings, within a range that is relevant for most practical applications considering, e.g., systems in the presence of thermal gradients, which will be analyzed elsewhere. The results of this analysis can be relevant to many other situations in which density- and temperature-dependent potentials arise from a CG process. In conclusion, our aim is not to provide an empirical model tailored to a single substance, but rather to establish a general framework from which GenDPDE parametrizations can be constructed for a broad class of relatively simple fluids. In this context, argon constitutes a natural first test case, since it is a condensed fluid with a well-characterized thermodynamic behavior, yet it does not exhibit the strong directional association or competing local structures that would obscure the assessment of the local thermodynamic model itself. More generally, the present study contributes to the methodology of CG molecular simulation by providing a statistically grounded framework for the parametrization of GenDPDE models with density- and temperature-dependent many-body interactions. In this way, it addresses the broader problem of connecting effective mesoscale particle models with target macroscopic EoS and local structural information.

The manuscript is organized as follows. In Section \ref{Sec-GenDPDE}, we briefly present the GenDPDE theoretical framework and numerical algorithm. In Section \ref{Sec-Thermo}, we introduce the particle LTh model for the description of condensed phases and derive the associated macroscopic EoS, expanding upon the importance of accounting for local structure effects, even at the mean-field level. We subsequently describe the procedure to consistently set the parameters of the mesoscopic model from macroscopic quantities. In Section \ref{Sec-Results}, we prove the accuracy of the proposed approach by analyzing the outcome of equilibrium GenDPDE simulations. We compare the numerical results, in terms of thermodynamic quantities, with data available from the NIST WebBook database \cite{NIST2024}. Moreover, we compare the radial distribution functions obtained from mesoscopic simulations with theoretical predictions using the Hypernetted Chain (HNC) approximation \cite{Hansen2006, Merabia2007}, discussing the accuracy and limitations of the latter. Finally, in Section \ref{Sec-Fin}, we summarize the main findings of this work.

\section{The \texorpdfstring{G\MakeLowercase{en}}{Gen}DPDE framework} \label{Sec-GenDPDE}

\subsection{Particle thermodynamics}

Within the GenDPDE framework, particles are regarded as mesoscopic objects which carry a given set of physical properties. If the physical system contains one single component, the state of the mesoparticle $i$ is described by its center-of-mass position $\br_i$, its total momentum $\bp_i$, its internal energy $u_i$, its mass $m_i$, and its volume $\mathcal{V}_i$. The internal energy $u_i$ is simply defined as the energy content of the non-resolved DoF, necessary for the energy conservation in the dynamic processes \cite{Avalos2019}. The mass of the mesoparticle depends on the degree of CG, $\phi$, which specifies the number of embedded physical constituents in each mesoparticle. Indicating the mass of these constituents by $m_w$, the mesoparticle mass will thus be $m_i=\phi \, m_w$. Throughout this article, $m_i$ is taken as constant and simply referred to as $m$. The volume of the mesoparticle is typically obtained in density-dependent potentials from a local estimate of the particle density, according to $\mathcal{V}_i^b \equiv 1/n_i^b$, with \cite{Pago2000, Warren2001, Zhao2021}
\begin{align}
n_i^b = \sum_{j \neq i} w(r_{ij}) \label{density}
\end{align}
where $r_{ij} = |\br_i-\br_j|$ and $w$ is a positive-definite, spherically symmetric, monotonically decreasing kernel, known as the {\em weighting function}, which vanishes when $r_{ij} \ge \Rc$, $\Rc$ being the so-called cutoff radius. This kernel is furthermore normalized such that its volume integral equals the unity. Here, the superscript $b$ is used to differentiate this {\em primitive} definition of the local density from the alternative definition introduced in Ref. \cite{Colella2025} and presented below. 

The density definition from Eq. \eqref{density} is widespread in DPD-like models and also in Smoothed Particle Hydrodynamics (SPH) \cite{Gingold1977,Lucy1977,Monaghan1992}. However, the evaluation of the particle volume from a local density calculated via a kernel $w$ with the aforementioned properties is responsible for several artifacts observed in SPH, such as the pairing instability \cite{Swegle1995,Price2012}, which fostered the application of the Wendland kernels to minimize its impact \cite{Dehnen2012}. When simulating liquids with GenDPDE, we face situations where the interparticle interactions dominate over thermal agitation. Under these conditions, we often observe the pairing instability in dynamic processes, accompanied by the formation of spurious local particle arrangements within the range $\Rc$, observable also in the pair distribution function $g(r)$ as shoulders in regions below the average interparticle distance, $l \simeq (1/\occ)^{1/3}$, with $\occ \equiv N/V$, with $N$ the total number of mesoparticles and $V$ the volume of the system. In Ref. \cite{Colella2025}, we demonstrate that this behavior is due to the inappropriate evaluation of the particle volume directly through Eq. \eqref{density} for finite $\Rc$. We show that such issues can be eliminated if an approximated {\em partition of the unity} approach \cite{Flekkoy1999} is followed, yielding an alternative expression for the particle volume that reads,
\begin{align}
    \mathcal{V}_i = \,\ &\frac{4 \pi}{3} \tRc^3 - 4 \pi \tRc^3 \sqrt{k_i} \bigg\{ (1-k_i)\arctan{\left(\frac{1}{\sqrt{k_i}}\right)} \nonumber \\
    &+ \sqrt{k_i} \left[ 1 - \ln{\left(\frac{k_i+1}{k_i}\right)} \right] \bigg\} \label{eq:1iv}
\end{align}
where $k_i = \frac{2\pi}{15} \tRc^3 n_i^b$ and $\tRc \equiv \Rc/f_{\text{cut}}$, with $f_{\text{cut}} > 1$ a scaling factor for the original cutoff radius that is tuned to obtain the desired value of $\mathcal{V}_i$. The advantage of this evaluation of the particle volume lies in the fact that it is ultimately related to the local density estimation $n_i^b$ (through $k_i$) and produces pairwise additive central forces, as in the standard DPD-like models. In what follows, we will consider the particle density $n_i$, defined from,
\begin{align}
    n_i \equiv \frac{1}{\mathcal{V}_i}  \label{newvol}
\end{align}
as the relevant variable related to the particle volume.

\subsection{Statistical interpretation: bare and dressed variables}

Given the description of the mesoparticles as property carriers, GenDPDE requires the definition of a particle LTh model, which allows us to define the particle pressure and temperature as functions of the state variables, namely, the internal energy $u$ and the local density $n$. A model including a particle variable composition has also been introduced \cite{Avalos2022,Lisal2022b,Colella2024}. We are implicitly assuming that the fluid particle is in thermal equilibrium, so that one can define a particle {\em bare} entropy function
\begin{align}
    \tS(u,n) \equiv k_B \ln g(u,n)   \label{entropy}
\end{align}
where $g(u,n)$ is the density of states of the non-resolved DoF \cite{Seifert2012} associated to a given particle. We implicitly assume no direct interaction between the unresolved DoF of neighbouring particles beyond that mediated by the resolved DoF.
%We are also implicitly assuming that there is no interaction between the non-resolved DoF of neighboring particles, other that through the resolved degrees of freedom, which is a reasonable assumption provided that the mesoparticle embeds a sufficient number of physical constituents, i.e., that the {\em level of CG} is sufficiently large, so that there are many DoF that are responsible for the internal state of the mesoparticle. 

Equation \eqref{entropy} allows us to define the local {\em intensive} variables. Effectively, the particle bare temperature $\tilde{\theta}$ is obtained in analogy with macroscopic thermodynamics \cite{Callen1985}, as,
\begin{align}
    \frac{1}{\tT} \equiv \left. \frac{\partial \tS}{\partial u}\right|_n   \label{bareT}
\end{align}
In the same way, the bare particle pressure follows from,
\begin{align}
    \frac{\tP}{\tT} \equiv \left. \frac{\partial \tS}{\partial \mathcal{V}}\right|_u   \label{bareP}
\end{align}
As $\tS(u,n)$ is a well behaved function, and $\tT > 0$ (it is strictly non-zero), the function $u(\tS,n)$ exists and the following relations also apply:
\begin{align}
    \tT =& \, \left. \frac{\partial u}{\partial \tS}\right|_n   \label{bareT2} \\
    \tP =& \, -\left. \frac{\partial u}{\partial \mathcal{V}}\right|_{\tS} = n^2 \left. \frac{\partial u}{\partial n}\right|_{\tS}   \label{bareP2}
\end{align}
%
%Analogously, the bare particle pressure follows from,
%
%\begin{align}
%    \tP = -\left. \frac{\partial u}{\partial \mathcal{V}}\right|_{\tS} = n^2 \left. \frac{\partial u}{\partial n}\right|_{\tS}   \label{bareP2}
%\end{align}
%
Given Eq. \eqref{entropy}, according to Einstein's theory of thermodynamic fluctuations \cite{Callen1985}, the probability distribution of the state of a system of $N$ mesoscopic particles in the canonical ensemble is given by,  
\begin{align}
P_\text{eq} (\tilde{\Gamma}) d\tilde{\Gamma} \propto e^{-\sum_i \left[\frac{\bp_i^2}{2m_i} +u_i -T \tilde{s}(u_i,n_i)	\right]/k_BT} d\tilde{\Gamma}	\label{barePeq}
\end{align}
where the state of the system is specified by the $7N$-dimensional vector $\tilde{\Gamma} = (\br_1,\br_2,\dots,\br_N, \\ \bp_1,\bp_2,\dots,\bp_N,u_1,u_2,\dots,u_N)$. In a mesoscopic system, Eq. \eqref{barePeq} is the fundamental equation, rather than the LTh description. In this expression, $\tilde{\Gamma}$ specifies that the internal energy $u$ is the independent fluctuating variable, together with the mesoparticle position and momentum. Introducing a change of independent variable by substituting $u$ with the {\em dressed} entropy $s$, defined from the relation,
\begin{align}
\tilde{s} = s+ k_B \ln \left( \Theta \, \frac{\partial s}{\partial u} \bigg|_n \right) \label{trans}
%= s + k_B \ln \left(\frac{\Theta}{\theta}\right)	
\end{align}
resulting from the absorption of the Jacobian of the transformation into the definition of the dressed entropy, we can write Eq. \eqref{barePeq} in terms of a different set of independent variables, namely, $\Gamma =  (\br_1,\br_2,\dots,\br_N,\bp_1,\bp_2,\dots,\bp_N,s_1,s_2,\dots,s_N)$. In Eq. \eqref{trans}, $\Theta$ is a scale of temperature defined later on. This transformation satisfies the condition that the probability distribution Eq.~\eqref{barePeq} should remain the same, i.e.,
\begin{align}
P_\text{eq} (\tilde{\Gamma}) d\tilde{\Gamma} = P_\text{eq} (\Gamma) d\Gamma  \label{equivalence}
\end{align}
Hence, the new probability density reads,
\begin{align}
P_\text{eq} (\Gamma) d\Gamma \propto e^{-\sum_i\left[\frac{\bp_i^2}{2m_i} +u(s_i,n_i) -T s_i \right]/k_BT} d\Gamma	\label{dressPeq}
\end{align}
We thus infer that the choice of the independent variable changes the LTh description, as the functions $\tS$ and $s$ do not have the same dependence on the energy. Therefore, a new set of independent intensive variables can also be defined from the dressed entropy, Eq. \eqref{trans}. Effectively, the dressed particle temperature and pressure read
\begin{align}
    \theta =& \, \left. \frac{\partial u}{\partial s}\right|_n   \label{dressT} \\
    \pi = & \, n^2 \left. \frac{\partial u}{\partial n}\right|_{s}   \label{dressP}
\end{align}
There are substantial differences between bare and dressed variables in their role as {\em estimators} of the corresponding macroscopic properties. It can be shown that \cite{Avalos2019}
\begin{align}
    \left\langle \frac{1}{\tilde{\theta}} \right\rangle = & \,\ \frac{1}{T}  \label{bareTav} \\[1mm] 
    \langle \theta \rangle =& \,\ T   \label{dressedTav}
\end{align}
where $T$ stands for the macroscopic {\em measurable} temperature of the reservoir. Notice, however, that $\langle \tT \rangle \neq T$ and, conversely, $\langle 1/\theta \rangle \neq 1/T$. For the macroscopic pressure $P$, the corresponding estimators approximately satisfy
\begin{align}
    \left\langle \frac{\tilde{\pi}}{\tilde{\theta}} \right\rangle \simeq& \,\ \frac{P}{T} - \bar{c}\,k_B = \frac{P^\text{ex}}{T}\label{barePav} \\[1mm]
    \langle \pi \rangle \simeq& \,\ P-\bar{c}\,k_B T = P^\text{ex}   \label{dressedPav}
\end{align}
as both $\tilde{\pi}$ and $\pi$ are related to the {\em excess} pressure of the ensemble of mesoparticles $P^\text{ex}$ \cite{Pago2001,Avalos2019}. Here, $\bar{c} = N/V$ is the system (bulk) number density, $V$ being the system total volume. Notice, however, that Eqs. \eqref{barePav} and \eqref{dressedPav} are strictly correct only in the limit of weakly-interacting systems, where correlations between particles can be neglected. In liquids, where particle correlations are important, the relationship between mesoscopic and macroscopic pressure is more complex and is influenced by the local structure of the system, as explained in more detail in Section \ref{Sec-Thermo}. In Refs. \cite{Avalos2019,Avalos2021}, we discuss the difference between bare and dressed variables in depth. In the following, we will formulate the LTh model in terms of dressed variables, for convenience, although transformations into bare variables will be indicated when needed.

\subsection{The GenDPDE algorithm}

The discrete Equations of Motion (EoM) for the dynamics of a GenDPDE system describe the time evolution of the resolved DoF, i.e., the mesoparticle position, momentum and internal energy, the latter obtained from total energy balance \cite{Avalos1997, Avalos2019}. These EoM read:
\begin{align}
     \br_{i}' =& \,\ \br_{i} + \frac{\bp_{i}}{m_i} \delta t \label{EoMr} \\
     \bp_{i}' =& \,\ \bp_{i} + \sum_{j \neq i} \bff_{ij}^{C} \delta t + \sum_{j \neq i} \bff_{ij}^{D} \delta t + \sum_{j \neq i} \delta \bp_{ij}^{R} \label{EoMp} \\
     u_{i}' =& \,\ u_i - \frac{1}{2}\sum_{j \neq i} \left( \frac{\bp_{i}}{m_i} - \frac{\bp_{j}}{m_j} \right) \cdot \bff_{ij}^{C} \delta t  \nonumber \\
                &- \frac{1}{2} \sum_{j \neq i} \left( \frac{\bp_{i}}{m_i} - \frac{\bp_{j}}{m_j} \right) \cdot \bff_{ij}^{D} \delta t  \nonumber \\
                &- \frac{1}{2} \sum_{j \neq i} \left( \frac{\bp_{i}}{m_i} - \frac{\bp_{j}}{m_j} \right) \cdot \delta \bp_{ij}^{R}  \nonumber \\
                &- \frac{1}{2m_i} \sum_{j \neq i}\sum_{l \neq i} \delta \bp_{ij}^{R} \cdot \delta \bp_{il}^{R}  \nonumber \\
                &+ \sum_{j \neq i} \dot{q}_{ij}  \delta t + \sum_{j \neq i} \delta u_{ij}^{R} \label{EoMu}
\end{align}
Here, the non-primed variables refer to a time instant $t$, while primed variables refer to the instant $t + \delta t$, with $\delta t$ the time step. This formulation highlights the causal nature of the stochastic algorithm, thus avoiding any possible ambiguity in its interpretation. In Eq. \eqref{EoMp}, the term $\bff_{ij}^{C}$ is the conservative force acting between a pair of particles, $i$ and $j$, and is defined as
\begin{align}
    \bff_{ij}^{C} =& \,\  - \frac{\partial u_i}{\partial n_i}\bigg|_{s_i} \frac{\partial n_i}{\partial \br_j} + \frac{\partial u_j}{\partial n_j}\bigg|_{s_j} \frac{\partial n_j}{\partial \br_i} \nonumber \\
    =& \,\  - \frac{\pi_i}{n_i^2} \frac{\partial n_i}{\partial \br_j} + \frac{\pi_j}{n_j^2} \frac{\partial n_j}{\partial \br_i}  \label{fc}
\end{align}
The last line of Eq. \eqref{fc} has been obtained using the definition Eq. \eqref{dressP}. Furthermore, considering Eq. \eqref{newvol}, the partial derivative of the density with respect to the position vector reads,
\begin{equation} \label{dnc1}
    \frac{\partial n_i}{\partial \br_j} = \zeta_i \frac{\partial n_i^b}{\partial \br_j}
\end{equation}
where
\begin{equation} \label{dnc2}
     \zeta_i \equiv \frac{\partial n_i}{\partial n_i^b} = \frac{\partial (1/\mathcal{V}_i)}{\partial n_i^b} = - \frac{1}{\mathcal{V}_i^2} \frac{\partial \mathcal{V}_i}{\partial k_i} \frac{\partial k_i}{\partial n_i^b}
\end{equation}
with
\begin{align}
    \frac{\partial \mathcal{V}_i}{\partial k_i} = &-4 \pi \tRc^3 \bigg[ \frac{3}{2} + \frac{1-3k_i}{2\sqrt{k_i}}\arctan{\left(\frac{1}{\sqrt{k_i}}\right)} \nonumber \\
    &-\ln{\left(\frac{k_i+1}{k_i}\right)}\bigg] \label{dVdk}
\end{align}
and 
\begin{equation} \label{dkdn0}
    \frac{\partial k_i}{\partial n_i^b} = \frac{2\pi}{15} \tRc^3
\end{equation}
in three dimensions. Thus, Eq. \eqref{fc} reads \cite{Avalos2019, Colella2025},
\begin{equation} \label{fcnew}
    \bff_{ij}^{C} = - \left( \frac{\pi_i}{n_i^2}\zeta_i + \frac{\pi_j}{n_j^2}\zeta_j \right) \frac{dw_{ij}}{dr_{ij}}\be_{ij}
\end{equation}
with $\be_{ij} = \br_{ij}/r_{ij}$ the separation-distance unit vector, and $w_{ij} \equiv w(r_{ij})$. The contribution $\bff_{ij}^{D}$ in Eq. \eqref{EoMp} represents the friction force as customarily defined in DPD-like methods, i.e.,
\begin{equation} \label{fd}
     \bff_{ij}^{D} = - \gamma_{ij} \left( \frac{\bp_i}{m_i} - \frac{\bp_j}{m_j} \right) \cdot \be_{ij} \be_{ij}
\end{equation}
where $\gamma_{ij}=\gamma \, \omega^p (r_{ij})$, with $\gamma$ the mesoscopic friction coefficient and $\omega^p (r_{ij})$ a kernel analogous to $w(r_{ij})$ but vanishing for $r_{ij} \ge \Rc^p$, with $\Rc^p$ the cut-off radius for this function. Notice however that, unlike $w(r_{ij})$, $\omega^p (r_{ij})$ may not be normalized. The last term in Eq. \eqref{EoMp}, $\delta \bp_{ij}^{R}$, represents the random contribution to the particle momentum, related to the dissipative term via the appropriate FD theorem \cite{Avalos2019},
\begin{equation} \label{fr}
     \delta \bp_{ij}^{R} = \sqrt{k_B (\theta_i+\theta_j) \gamma_{ij}} \,\ \Omega_{ij}^p \be_{ij} \delta t^{1/2}
\end{equation}
with $\Omega_{ij}^p$ a normalized Gaussian number satisfying
\begin{align}
    \langle \Omega_{ij}^{p} \rangle &= 0 \label{random1} \\
    \langle \Omega_{ij}^{p} (t) \Omega_{kl}^{p} (t') \rangle &= \left( \delta_{ik}\delta_{jl} - \delta_{il}\delta_{jk} \right) \delta_{tt'} \label{random2}
\end{align}
Here, $\delta_{ij}$ stands for the Kronecker delta, so that $\delta_{ij}=1$ if $i=j$, and $\delta_{ij}=0$ otherwise. On the other hand, $\delta_{tt'}=1$ only when $t$ and $t'$ refer to the same instant (within the time discretization), being equal to 0 otherwise. The term $\dot{q}_{ij}$ appearing in Eq. \eqref{EoMu} is the heat flux between particles $i$ and $j$, defined as
\begin{equation} \label{qdot}
    \dot{q}_{ij} = - \kappa_{ij} \left( \frac{1}{\tilde{\theta}_j} - \frac{1}{\tilde{\theta}_i} \right)
\end{equation}
where $\kappa_{ij}=\kappa \, \omega^u (r_{ij})$, with $\kappa$ the mesoscopic thermal conductivity and $\omega^u (r_{ij})$ a kernel analogous to $\omega^p (r_{ij})$ and vanishing for $r_{ij} \ge \Rc^u$, with $\Rc^u$ the associated cut-off length. Notice that, here, the heat flux has been written in terms of the bare temperatures, following the traditional derivation of Ref. \cite{Avalos2019}. A reformulation of this quantity in terms of dressed variables can be introduced by applying the suitable transformation from $\tilde{\theta}$ to $\theta$, and redefining the mesoscopic thermal conductivity accordingly (see Ref. \cite{Avalos2021} for a detailed discussion). Finally, the last term in Eq. \eqref{EoMu}, $\delta u_{ij}^{R}$, is the random transferred heat, related to the heat flux through a FD theorem \cite{Avalos2019} and reading,
\begin{equation} \label{qr}
     \delta u_{ij}^{R} = \sqrt{2 k_B \kappa_{ij}} \,\ \Omega_{ij}^u \delta t^{1/2}
\end{equation}
with $\Omega_{ij}^u$ also a normalized Gaussian number with properties Eqs. \eqref{random1} and \eqref{random2}.

\section{Thermodynamics} \label{Sec-Thermo}

The ensemble of mesoparticles with a given LTh model has a collective behavior described by a set of EoS relating the macroscopic variables, such as the internal energy $U$ or the pressure $P$, with the system temperature $T$, volume $V$, and number of mesoparticles $N$ (as we consider here homogeneous systems of one single component). In addition, one must also consider the degree of CG, $\phi$, as a parameter, since the total number of physical constituents $\mathcal{N}$ is related to the number of mesoparticles $N$ from the decimation relation $\mathcal{N} = \phi N$. In this section, we first derive the explicit LTh model to be considered in our work, and, subsequently, the macroscopic EoS for the pressure and the internal energy in terms of the model parameters. These equations, when compared with experimental values available in databases \cite{NIST2024}, will be used to set the model parameters for the analysis of a given physical system.

\subsection{Local thermodynamic model for single-component systems} \label{LThmodel}
%\subsection{Local thermodynamic model for nearly homogeneous systems} \label{LThmodel}

In this work, we propose a model that is valid for describing liquids and supercritical fluids subject to small deviations in temperature, density, and pressure, of the order of a few percent, from a given reference state. Under this assumption, the LTh can be set by the knowledge of a few coefficients, namely, the thermal expansion coefficient, $\alpha$, the isothermal compressibility, $\kappa_T$, and the heat capacity at constant volume, $C_V$, of the mesoparticle, all of them evaluated at the reference state. More sophisticated models could also be introduced (see, e.g., Refs. \cite{Avalos2019,Avalos2021}), with which phenomena, such as, e.g., phase separation, can be reproduced. 

In what follows, we consider the particle Helmholtz free energy $f(\theta,n)$ as the fundamental thermodynamic potential containing all the information about the LTh of the mesoparticle. Notice that $\theta$ is the particle dressed temperature, Eq. \eqref{dressT}, and $n$ is the local density calculated from Eq. \eqref{newvol}. Defining $f(\theta, n)$ guarantees thermodynamic consistency of the derived properties. We demand that the LTh model for the particle satisfy several important conditions. The first one is that the entropy must be bound at $\theta \rightarrow 0$, but also that 
\begin{align}
\left. \frac{\partial s}{\partial \theta} \right|_n \to 0 \label{cond1}
\end{align}
as $\theta \to 0$, in agreement with the Third Law of Thermodynamics. This also implies that the heat capacity $C(\theta,n) \equiv \frac{\partial u}{\partial \theta}\big|_n$ satisfies the condition
\begin{align}
\frac{C(\theta \to 0)}{\theta} \to 0	\label{3rdlaw}
\end{align}
as $\theta \to 0$. Moreover, the internal energy must also be bound in this limit. 

\subsubsection{Simple Equations of State for condensed phases}

Let us define the LTh reference state through a particle pressure $\pi_0$ and a temperature $\theta_0$. The isothermal compressibility of the mesoparticle at this particular state is $\kappa_T$. In the liquid phase, we expect the compressibility to be very small, meaning that small variations in volume will produce large variations in pressure, according to the definition, 
\begin{align}
\kappa_T \equiv - \left. \frac{1}{\mathcal{V}} \frac{\partial \mathcal{V}}{\partial \pi} \right|_\theta = \left. \frac{1}{n} \frac{\partial n}{\partial \pi} \right|_\theta \label{k}
\end{align}
Integrating Eq. \eqref{k}, we obtain,
\begin{align}
n(\pi,\theta) =& \, n(\pi_0,\theta) e^{\kappa_T(\theta) (\pi-\pi_0)} \equiv \nonumber \\
\equiv& \, n_0(\theta) e^{\kappa_T(\theta) (\pi-\pi_0)} 	\label{simplex0}
\end{align}
For simplicity, from here on we will ignore the possible temperature dependence of $\kappa_T$, which will thus be considered as constant in the range of interest of temperatures and pressures. The accuracy of this approximation will be tested against experimental data in Sec. \ref{Sec-Results}. We can also introduce the definition of the thermal expansion coefficient for the mesoparticle, 
\begin{align}
\alpha \equiv -\frac{1}{n} \left. \frac{\partial n}{\partial \theta} \right|_{\pi} = -\frac{1}{n_0} \left. \frac{\partial n_0}{\partial \theta} \right|_{\pi_0}  \label{alpha}
\end{align}
Integrating Eq. \eqref{alpha} from a reference temperature $\theta_0$, we obtain,
\begin{align}
n_0(\theta) =& \, n_0(\theta_0) e^{-\alpha(\theta-\theta_0)} \equiv \nonumber \\
\equiv& \, n_{00} e^{-\alpha(\theta-\theta_0)}	\label{simplex00}
\end{align}
where $n_{00} \equiv n(\pi_0,\theta_0)$. Therefore,
\begin{align}
n(\pi,\theta) = n_{00} e^{-\alpha(\theta-\theta_0) }e^{\kappa_T (\pi-\pi_0)} \label{Gibbs}
\end{align}
To determine the Helmholtz free energy $f$, we require that $n$ and $\theta$ be the independent variables. Thus, we reformulate Eq. \eqref{Gibbs} as,
\begin{align}
\pi(n,\theta) = \pi_{00} + \frac{\alpha}{\kappa_T} (\theta-\theta_0) + \frac{1}{\kappa_T} \ln \frac{n}{n_{00}} \label{simplex}  
\end{align}
where $\pi_{00} \equiv \pi(n_{00},\theta_0)$. This is the first Equation of State (EoS) for the particle LTh model. Notice that, in this expression, the pressure is a linear function of the particle temperature. Equation \eqref{simplex} can be integrated to obtain the Helmholtz free energy, yielding,
\begin{align}
f = & - \frac{\pi_{00}}{n} - \frac{\alpha}{n \kappa_T}(\theta - \theta_0) -\frac{1}{n\kappa_T} \left( \ln \frac{n}{n_{00}} +1 \right) + \Phi(\theta) \label{f}
\end{align}
where $\Phi(\theta)$ is a yet unknown function of the particle temperature only.
%
%\begin{align}
%f = & - \frac{\pi_{00}}{n} - \frac{\alpha}{n \kappa_T}(\theta - \theta_0) -\frac{1}{n\kappa_T} \left( \ln \frac{n}{n_{00}} +1 \right) - \nonumber \\
%&  -\frac{1}{2} C_V \theta \ln (1+\lambda^2 \theta^2) \label{fsimplex}
%\end{align}
%
To set the functional form of $\Phi$, let us first calculate the particle entropy,
\begin{align}
s \equiv -\left.\frac{\partial f}{\partial \theta}\right|_n = \frac{\alpha}{n\kappa_T} - \Phi'(\theta)		\label{s}
\end{align}
The internal energy $u = f+\theta s$ then reads,
\begin{align}
u =  - \frac{\pi_{00}}{n} + \frac{\alpha}{n \kappa_T}\theta_0 -\frac{1}{n\kappa_T} \left( \ln \frac{n}{n_{00}} +1 \right) + u_\text{id}(\theta)   \label{u}
\end{align}
where we have defined $u_\text{id} \equiv \Phi -\theta \Phi'$. Due to the linear dependence of the pressure with the temperature, the internal energy can be split into one term that solely depends on $n$ and another term that depends only on $\theta$. In the low-temperature limit, the Debye model for the heat capacity \cite{Debye1912} reveals that $C(\theta \to 0) = u_\text{id}' (\theta \to 0) \propto \theta^3$. Hence, we can write,
\begin{align}
u_\text{id}(\theta \to 0) \to C_V\frac{\theta^4}{\Theta^3} \label{CVDebye}
\end{align}
where $\Theta$ is the crossover temperature scale (analogous to the Debye temperature) already introduced in Eq. \eqref{trans}, and $C_V$ is a constant with the dimensions of the heat capacity. For simplicity, we also demand that the internal energy $u_\text{id}$ tend to the classical ideal-gas limit for $\theta \to \infty$, i.e.,
\begin{align}
u_\text{id}(\theta \to \infty) \to C_V \theta \label{CVideal}
\end{align}
Therefore, one can interpolate between the two limits with a function of the form,
\begin{align}
u_\text{id}(\theta) = C_V \theta \frac{\theta^3/\Theta^3}{1+\theta^3/\Theta^3}	\label{uid}
\end{align}
The EoS for the particle energy then reads,
\begin{align}
u =& \, - \frac{\pi_{00}}{n} + \frac{\alpha}{n \kappa_T}\theta_0 - \frac{1}{n\kappa_T} \left( \ln \frac{n}{n_{00}} +1 \right) + \nonumber \\
&+ \, C_V \theta \frac{\theta^3/\Theta^3}{1+\theta^3/\Theta^3}  \label{u2}
\end{align}
Next, we integrate the differential equation $u_\text{id} = \Phi-\theta \Phi'$ to obtain the functional form of $\Phi$, i.e.,
\begin{align}
\Phi(\theta) = - \frac{1}{3} C_V \theta \ln \left(1+\frac{\theta^3}{\Theta^3}\right)   \label{g}
\end{align}
This last expression together with Eq. \eqref{s} allows us to obtain the complete expression for the particle entropy, yielding,
\begin{align}
s = & \, \frac{\alpha}{n\kappa_T} + \frac{1}{3} C_V \ln \left(1+\frac{\theta^3}{\Theta^3}\right) + C_V \frac{\theta^3/\Theta^3}{1+\theta^3/\Theta^3}\label{ss}
\end{align}
On the other hand, from Eqs. \eqref{f} and \eqref{g}, we finally arrive at the explicit form of the particle Helmholtz free energy,
\begin{align}
f = & - \frac{\pi_{00}}{n} - \frac{\alpha}{n \kappa_T}(\theta - \theta_0) -\frac{1}{n\kappa_T} \left( \ln \frac{n}{n_{00}} +1 \right) - \nonumber \\
&  - \frac{1}{3} C_V \theta \ln \left(1+\frac{\theta^3}{\Theta^3}\right)
\label{fexplicit}
\end{align}
All the relevant particle LTh properties can be derived from this last equation. The coupling between the temperature and the density in Eq. \eqref{fexplicit} leads to a temperature-dependent potential force for the particle, which permits the analysis of non-isothermal situations in a natural and consistent way.

To end this section, we construct the function $u(s,n)$, which is required for the statistical mechanical analysis of the model, presented below. The previous expressions can be greatly simplified for fluids at moderately high temperatures, as is the case, e.g., at ambient conditions. In this limit, i.e., when $\theta \gg \Theta$, the third law of Thermodynamics is not relevant. From Eq. \eqref{u2} we can thus write,
\begin{align}
    u = C_V \theta + {\cal V}(n)
\end{align}
from which one can obtain a closed expression for the particle temperature in terms of the internal energy,
\begin{align}
    \theta = \frac{u-{{\cal V}}(n)}{C_V}  \label{temp}
\end{align}
Here, we have introduced the potential ${\cal V}(n)$, which gathers all the terms in the internal energy that are independent from the particle temperature,
\begin{align}
    {\cal V}(n) \equiv - \frac{\pi_{00}}{n} + \frac{\alpha}{n \kappa_T}\theta_0 - & \frac{1}{n\kappa_T} \left( \ln \frac{n}{n_{00}} +1 \right) \label{pot}
\end{align}
Next, replacing Eq. \eqref{temp} in Eq. \eqref{ss} in the high-temperature limit we find,
\begin{align}
s = & \frac{\alpha}{n\kappa_T} + C_V \ln \left(\frac{u-{\cal V}(n)}{C_V \Theta}\right) + C_V \label{su}
\end{align}
From this expression we can readily obtain,
\begin{align}
    u = & \, {\cal V}(n) + C_V\Theta \exp \left[\left(s-C_V-\frac{\alpha}{n\kappa_T} \right)/C_V \right] \nonumber \\
    = & \, {\cal V}(n) + C_V \Theta \, e^{s/C_V} [\psi(n)]^{k_B/C_V}   \label{us}
\end{align}
where we have introduced the function
\begin{align}
\psi(n) \equiv e^{-\left(C_V+\frac{\alpha}{n \kappa_T}\right)/k_B}	\label{defpsi}
\end{align}
to cast the present LTh model in the form of general density-dependent models with linear temperature dependence \cite{Avalos2019}. With this general notation, the particle pressure reads, 
\begin{align}
\pi =& \, k_B \theta \frac{n^2 \, \psi'(n)}{\psi(n)} + n^2 \, {\cal V}'(n) = \nonumber \\
=& \, \frac{\alpha}{\kappa_T} \theta + n^2 \, {\cal V}'(n) 	\label{genpi}
\end{align}

Finally, using Eq.~\eqref{trans}, we can establish a relationship between bare and dressed variables. For the temperature, we find that, 
\begin{align}
\frac{1}{\tilde{\theta}} = \frac{1}{\theta} \left(1- \frac{k_B}{C(\theta)}	\right) \simeq \frac{1}{\theta} \left(1- \frac{k_B}{C_V}	\right) \label{equivT}
\end{align}
where we have introduced the heat capacity as
\begin{align}
 C(\theta) = C_V \left[ \frac{4 (\theta/\Theta)^3 + (\theta/\Theta)^6}{(1+(\theta/\Theta)^3)^2} \right]  \label{heatcapacity}
 %C_V \frac{\theta^2}{\Theta^2} \frac{1}{1+\frac{\theta^2}{\Theta^2}} \left(1+2 \frac{1}{1+\frac{\theta^2}{\Theta^2}} \right)
\end{align}
which becomes $C(\theta \gg \Theta) \to C_V$ in the high-temperature limit, leading to the last equality in Eq. \eqref{equivT}. For the pressure, since $\frac{\tilde{\pi}}{\tilde{\theta}n^2} =- \left. \frac{\partial \tilde{s}}{\partial n}\right|_u$, we can write,
\begin{align}
\frac{\tilde{\pi}}{\tilde{\theta}n^2} = \frac{\pi}{\theta n^2} + \frac{k_B}{\theta}\left. \frac{\partial \theta}{\partial n} \right|_u
\end{align}
After some algebra, in the high-temperature limit we obtain,
\begin{align}
\tP = \frac{\pi}{1-k_B/C_V} - \frac{k_B/C_V}{1-k_B/C_V} n^2 \mathcal{V}'(n)
\end{align}
% CHECKED: WRONG SIGN ONLY
%
Notice that the differences between bare and dressed variables scale with the size of the energy fluctuations, namely, $\mathcal{O}(k_B/C_V)$, and become negligible for very large levels of CG.

\subsection{Partition function}

Let us consider the partition function of the system, namely,
\begin{align}
Q_N(V,T) = &\frac{1}{\Lambda^{3N} \varepsilon^N \,N!} \int d\br_1 \dots d\br_N \, du_1 \dots du_N \times \nonumber \\
& \hspace{2.5cm} \times  e^{-\sum_i \tilde{\cal F}_i/k_BT}	\label{partition}
\end{align}
where $\tilde{{\cal F}}_i$ is given by,
\begin{align}
\tilde{\cal F}_i = u_i - T\tilde{s}(u_i,n_i) \label{free}
\end{align}
Notice that we have implicitly used Eq. \eqref{entropy} to replace the integration over the internal degrees of freedom by the integration over the energy stored for every particle. 
The factor $1/\varepsilon$ per particle stands for the measure of the separation between the energy levels of the non-resolved DoF, so that the summation over states can be transformed into an integral over the energy, $\sum_r \dots \simeq \int (du/\varepsilon) \dots$. The de Broglie thermal wavelength is $\Lambda \equiv \sqrt{h^2/2 \pi m kT}$, and appears after integration of the resolved momenta. Notice that the de Broglie wavelength depends on the mesoparticle mass $m$ and, therefore, on the degree of CG. Moreover, due to the additivity of the particle kinetic energy, we can factorize this integration for each particle.

The expression in Eq. \eqref{partition} is given in terms of bare variables. However, it can be consistently rewritten in terms of dressed variables, without loss of generality, reading,
\begin{align}\label{partition2}
Q_N(V,T) = &\frac{\Theta^N}{\Lambda^{3N} \varepsilon^N \,N!} \int d\br_1 \dots d\br_N \, ds_1 \dots ds_N \times \nonumber \\
& \hspace{2.5cm} \times  e^{-\sum_i {\cal F}_i/k_BT}
\end{align}
where
\begin{align}
 {\cal F}_i = u(s_i,n_i) - Ts_i \label{uands}
\end{align}
with $s_i$ given by Eq. \eqref{trans}. The temperature scale $\Theta$ ({\em cf}. \eqref{trans} and \eqref{CVDebye}) can be explicitly defined from the energy scale as
\begin{align}
    \Theta \equiv \frac{\varepsilon}{k_B}   \label{Debye}
\end{align}
and it is analogous to the Debye temperature in solids \cite{Callen1985}. The partition function is related to the macroscopic Helmholtz free energy, $F$, through a {\em fundamental relationship} that contains all the relevant thermodynamic information of the system. In the canonical ensemble, this relationship reads,
\begin{align}
F = -k_BT \ln Q_N	\label{helm}
\end{align}

For the particular case under analysis, the linearity of $u_i$ in $\theta_i$ allows us to analytically integrate the energy contribution in the partition function Eq. \eqref{partition2}. Using Eq.~\eqref{us} and the factorization of the exponential, the integration with respect to $s$ can be readily performed independently for every particle (see Appendix \ref{App-QN}), yielding ({\em cf.} Eq. \eqref{sint}),
\begin{align}\label{partition3}
Q_N(V,T) =& \, \frac{1}{\Lambda^{3N} \varepsilon^N \,N!} \int d\br_1 \dots d\br_N \times  \\
 &\times \prod_{i=1}^N \left[\frac{C_V \Theta}{\psi(n_i)} \left(\frac{k_BT}{C_V \Theta}	\right)^{\frac{C_V}{k_B}} e^{-\frac{{\cal V}(n_i)}{k_BT}} \Gamma \left( \frac{C_V}{k_B} \right)	\right] \nonumber
\end{align}
All the terms that do not depend on the particle density $n_i$ can be taken out of the integral, leading to a factorization of the form ({\em cf.} Eq.~\eqref{factorapp})
\begin{equation}
Q_N(V,T) = [q(T)]^N \, [Q^{id}(V,T)]^N \, Q_N^{ex}(V,T) \label{factor}
\end{equation}
where $q(T)$ is a function of the temperature and the parameters of the model only ({\em cf.} Eq.~\eqref{qapp}), $Q^{id}$ is the classical one-particle ideal-gas partition function ({\em cf.} Eq.~\eqref{idealgas}), and the excess contribution $Q_N^{ex}$ ({\em cf.} Eq.~\eqref{partitionapp6}) gathers all the dependencies on the particle positions, i.e.,
\begin{equation}
Q_N^{ex}(V,T) \equiv \frac{1}{V^N} \int d\br_1 \dots d\br_N \, e^{-\sum_i \left[\frac{{\cal V}(n_i)}{k_BT}+\ln \psi(n_i)\right]} \label{partition6}
\end{equation}
This last integration is reminiscent of the configurational integrals in standard statistical mechanics of molecular systems, which are solved via Monte Carlo simulations or via suitable approximations, to obtain the properties of the system. 

In comparison with the standard DPD model, the GenDPDE LTh model incorporates into the ideal part the effect of the internal DoF through $q(T)$. Moreover, the configurational integral takes a particular form in which the effective potential 
\begin{align}
    \mathcal{W}(T,n) \equiv \mathcal{V}(n) +k_BT \ln \psi(n) \label{efpot}
\end{align}
contains an additional term that explicitly depends on the temperature of the reservoir, due to the temperature-dependence of the force. Notice that the differentiation of Eq. \eqref{efpot} with respect to $n$ is related to the particle pressure at the reservoir temperature and, therefore, to the force between mesoparticles, i.e., 
\begin{align}
\frac{\partial {\cal W}}{\partial n} = \frac{\pi(T,n)}{n^2}    \label{difw}
\end{align}

Aiming at obtaining the macroscopic thermodynamic properties of the model, Eq. \eqref{partition6} needs to be solved. However, the many-body nature of the density-dependent potential $\mathcal{W}$ introduces several complications. Notice that, for potentials that depend parametrically on the global density, inconsistencies arise in the derivation of the EoS of the system following different routes \cite{Louis2002, Tejero2009}. Since the GenDPDE potential depends on a {\em local} density $n$, rather than on a global one, it is free from this type of conundrums. However, as the mesoparticles strongly interact through $\mathcal{W}$, the effects of the local structure cannot be ignored by using, e.g., a Random Phase Approximation \cite{Merabia2007}, which assumes that the dominant contribution is due to a homogeneous average density $N/V$. As a consequence, interparticle correlations need to be accounted for when calculating the thermodynamic properties of the system, even at the lowest level of approximation. 

\subsubsection{Considerations on the local density}

The interpretation of the local density requires some discussion. We recall that $n_i$ is a measure of the $i^{\text{th}}$ particle volume, according to Eqs. \eqref{eq:1iv} and \eqref{newvol}. The input for the latter is the primitive density $n_i^b$, which is obtained from neighboring particles' positions, in view of Eq. \eqref{density}. The instantaneous value of $n_i^b$, denoted by $\hat{n}_i^b$ is given by, 
\begin{align}
\hat{n}_i^b = & \,\  \hat{n}^b(\br_i) = \int d\br' \, w(|\br_i-\br'|) \sum_{j \neq i} \delta (\br'-\br_j) = \nonumber \\
= &  \int d\br' \, w(|\br_i-\br'|) \hat{c}^*(\br')
\end{align}
where the asterisk in $\hat{c}^*$ indicates that this quantity is a {\em conditional} density, due to the exclusion of  particle $i$. Hence, the instantaneous field $\hat{c}^*(\br)$ is still implicitly dependent on the position of the $i^\text{th}$ particle. In contrast, $\hat{c}(\br) \equiv \sum_i \delta (\br-\br_i)$ is the instantaneous {\em unconditioned} particle density. Thus, let us consider the following equilibrium average (denoted by $\langle \dots \rangle$) 
\begin{align}
 \langle \hat{c}(\br) \, \hat{n}^b(\br) \rangle 
= & \int d\br' \, w(|\br-\br'|) \langle \hat{c}(\br)\hat{c}^*(\br') \rangle
\end{align}
We can further write $\langle \hat{c}(\br)\hat{c}^*(\br')\rangle \equiv c(\br)c(\br') g(\br'|\br)$ \cite{Hansen2006}, with $g(\br'|\br)$ the pair distribution function. The chosen notation is intended to stress the fact that $g$ is related to the conditional probability of finding a particle at $\br'$ provided that there is a particle at $\br$. Hence, we can write,
\begin{align}
\langle \hat{c}(\br)  \hat{n}^b(\br) \rangle = \int d\br' \, w(|\br-\br'|) c(\br)c(\br') g(\br'|\br)
\end{align}
From this expression, we can introduce the average local density field $n^b(\br)$ as,
\begin{align}
n^b(\br) \equiv & \int d\br' \, w(|\br-\br'|) c(\br') g(\br'|\br) \label{local}
\end{align}
For a homogeneous, translationally invariant system, Eq. \eqref{local} simplifies to,
\begin{align}
n^b  = & \,\ \occ \int d\br' \, w(r') g(\br') \label{local2}
\end{align}
where $\overline{c}$ is the bulk density of the system. 

\subsubsection{Expansion of the potential in density fluctuations}

Having defined the mean density field, we now introduce the perturbation variable,
\begin{align}
    \hat{n}_i^b = n^b(\br_i) + \delta \hat{n}_i^b \label{perturb}
\end{align}
%
%where $\hat{n}_i$ is the instantaneous value of the local density $n_i$ as given in Eq. \eqref{newvol}, while 
where $n^b(\br_i)$ is the average density field given in Eq. \eqref{local}, using a predetermined pair distribution function $g_\text{ref}(\br'|\br)$. At this point, the function $g_\text{ref}(\br'|\br)$ is arbitrary as it simply defines the reference state for the expansion of the potential in powers of the deviation from such state. 

It is very important to realize that the size of the density fluctuations is related to the cutoff distance used to measure the local density, which also strongly affects the magnitude of the interparticle forces. Effectively, to obtain a rough estimation of the size of $\delta n$, let us consider a sphere of radius $\Rc$ that is randomly filled with particles, which otherwise fill the volume with a homogeneous density $\occ$. Thus, the probability of finding one arbitrary particle within the sphere surrounding particle $i$, say, is proportional to $\Rc^3/V$, and can be approximated by a Poisson distribution, $\mathbb{P}(\nu) = e^{-x} x^\nu/\nu!$, where $\nu$ is the number of neighbors and $x \propto \occ \Rc^3$ is its expected average value. Then, the fluctuations in the number of neighbors can be readily calculated, yielding $\big\langle \delta \nu^2 \big\rangle = x \propto \occ \Rc^3$. Hence, 
\begin{align}
    \frac{\big\langle \delta \nu^2 \big\rangle }{\big\langle \nu^2 \big\rangle } & \propto \frac{1}{\occ \Rc^3} \to 0 \;\; \mbox{as} \;\; \Rc \to \infty 
\end{align}
Therefore, when one particle moves in the sea of neighboring particles, it finds density variations of this order, which implies that the forces become harsher as $\Rc$ decreases, and actually disappear if $\Rc^3$ becomes of the order of the total volume, where the particle always finds a constant number of neighbors in its motion. As a consequence, as long as $\langle \delta \nu^2 \rangle/\langle \nu^2 \rangle \propto \Rc^{-3} \ll 1$, one can approximate,
\begin{align}
    \mathcal{W}(\hat{n}_i) =  \mathcal{W}(n(\br_i)) + \left.\frac{\partial {\cal W}}{\partial \hat{n}} \frac{\partial \hat{n}}{\partial \hat{n}^b}\right|_{n^b(\br_i)} & (\hat{n}_i^b-n^b(\br_i)) \, + \nonumber \\
    &+ \mathcal{O}(\delta \hat{n}_i^b)^2\label{Wexpansion}  
\end{align}
where we have exploited the fact that $\hat{n}_i = n(\hat{n}_i^b)$ is a function of $\hat{n}_i^b$ alone. In this expression, $n(\br_i) \equiv n(n^b(\br_i))$.
%It is also convenient to introduce here the factor
%
%\begin{align}
%   \zeta_i = \zeta(\hat{n}_i^b) \equiv \frac{\partial \hat{n}_i}{\partial \hat{n}_i^b} =  -\hat{n}_i^2 \frac{\partial \mathcal{V}_i}{\partial \hat{n}_i^b}
%\end{align}
%
%that indicates the variation of the actual local density when the primitive density $\hat{n}_i^b$ varies due to changes in particle positions.
Thus, the right-hand side of Eq. \eqref{Wexpansion} is divided into a one-particle contribution, due to the interaction with the average mean field, plus a pairwise contribution, namely,
%
%\begin{align}
%     \mathcal{W}(\hat{n}_i) \simeq& \, \mathcal{W}(n(\br_i)) - [\mathcal{W}_{n}\zeta_i]_{n^b(\br_i)} \, n^b(\br_i)+ \nonumber \\
%    & +  [\mathcal{W}_{n}\zeta_i]_{n^b(\br_i)} \,\sum_{j \neq i} w(\br_i-\br_j) \label{Wexpansion2}
%\end{align}
%
\begin{align}
    \mathcal{W}(\hat{n}_i) \simeq& \, [\mathcal{W}]_{n^b(\br_i)} - [\mathcal{W}_{n}]_{n^b(\br_i)} \, n^b(\br_i)+ \nonumber \\
   & +  [\mathcal{W}_{n}]_{n^b(\br_i)} \,\sum_{j \neq i} w(|\br_i-\br_j|) \label{Wexpansion2}
\end{align}
where we have defined, for the sake of simplicity, 
\begin{align}
 [\mathcal{W}]_{n^b(\br_i)} \equiv & \, \mathcal{W}(n(\br_i)) \label{W} \\
 [\mathcal{W}_{n}]_{n^b(\br_i)} \equiv & \left.\, \frac{\partial {\cal W}}{\partial n} \frac{\partial n}{\partial \hat{n}^b}\right|_{n^b(\br_i)} = \left.\frac{\pi}{\hat{n}^2} \zeta 
    \right|_{n^b(\br_i)}    \label{Wn}
\end{align}
The choice of this particular notation emphasizes that the expansion is made in terms of $n_i^b$ rather than of the corrected $n_i$. Thus, $n^b(\br_i)$, $[\mathcal{W}]_{n^b(\br_i)}$, as well as $[\mathcal{W}_{n}]_{n^b(\br_i)}$ are constant fields, independent of the instantaneous particle positions.

\subsection{Macroscopic Equations of State}

From the above analysis, we can derive the macroscopic EoS consistent with the approximations introduced so far. These EoS will produce the functional form of the system pressure $P$ and internal energy $U$ in terms of the mesoscopic parameters, which will ultimately allow us to set the appropriate parametrization of the model aiming to describe a given physical system close to some specified reference state.

\subsubsection{Pressure EoS}

To derive the equation for the pressure, we use the so-called energetic route, to prove the internal consistency of the statistical mechanical analysis \cite{Merabia2007}. Effectively, upon differentiation of Eq. \eqref{helm} with respect to the system volume, one finds, 
\begin{align}
    P =& \, \occ \, k_BT - \left\langle \frac{1}{3V} \sum_{i, j<i} \bigg(\frac{\pi_i}{n_i^2} \zeta_i +\frac{\pi_j}{n_j^2} \zeta_j \bigg) r_{ij} w'(r_{ij}) \right\rangle \equiv \nonumber \\
    \equiv& \,P^{\text{id}} + P^{\text{ex}} \label{virial} 
\end{align}
where $P^{\text{id}} \equiv \occ \, k_BT$ is the ideal-gas contribution to the system pressure due to the motion of the mesoparticles, whereas $P^{\text{ex}}$ represents the {\em excess} pressure, related to the particle pressure (see Appendix \ref{App-Merabia}). Using the same notation as in Eq. \eqref{Wn}, we can rewrite the excess pressure as,
\begin{align}
    P^{\text{ex}} =& \, -  \frac{1}{6V} \left\langle \sum_{i, j \neq i} \bigg([\mathcal{W}_n]_{\hat{n}_i^b} + [\mathcal{W}_n]_{\hat{n}_j^b} \bigg) r_{ij} w'(r_{ij}) \right\rangle \label{Pex}
\end{align}
where $[\mathcal{W}_n]_{\hat{n}_i^b}$ and $[\mathcal{W}_n]_{\hat{n}_j^b}$ are functions of the instantaneous primitive local density. Notice the change in the summation limits. Analogously to Eq. \eqref{Wexpansion2}, we can thus propose an expansion of these terms around the average density field Eq. \eqref{local}, yielding,
\begin{align}
    [\mathcal{W}_n]_{\hat{n}_i^b} \simeq & \,\ [\mathcal{W}_n]_{n(\br_i)} - [\mathcal{W}_{nn}]_{n^b(\br_i)} n^b(\br_i) \,\ + \nonumber \\ 
    &+ [\mathcal{W}_{nn}]_{n^b(\br_i)} \sum_{k \neq i}w(|\br_i-\br_k|) \label{Wnexp}
\end{align}
and analogously for the $j^{\text{th}}$ contribution. Here, we have defined,
\begin{align}
    [\mathcal{W}_{nn}]_{n^b(\br_i)} \equiv & \,\ \left. \frac{\partial \mathcal{W}_n}{\partial n} \frac{\partial n}{\partial n^b} \right|_{n^b(\br_i)} = \nonumber \\
    = & \,\ \bigg[ \frac{\zeta}{n^2} \bigg( \frac{\zeta}{\kappa_T \, n} + \pi\zeta_n \bigg) - \frac{2}{n^3} \pi \zeta^2 \bigg]_{n^b(\br_i)} \label{Wnn}
\end{align}
where $\zeta_n \equiv \frac{d\zeta}{d\hat{n}^b}$, and the definition of the isothermal compressibility Eq. \eqref{k} has been used. Next, introducing the instantaneous fields. After some algebra, one arrives at (see Appendix \ref{App-Merabia}),
%
%\begin{widetext}
\begin{align}
    P^{\text{ex}} =& - \frac{1}{3V} \int d\br d\br' [\mathcal{W}_n]_{n^b(\br)} |\br-\br'| w'(|\br-\br'|) c(\br)c(\br')g(\br'|\br) \,\ + \nonumber \\
    & - \frac{1}{3V} \int d\br d\br'[\mathcal{W}_{nn}]_{n^b(\br)} |\br-\br'| w(|\br-\br'|) w'(|\br-\br'|) c(\br)c(\br')g(\br'|\br) \,\ + \nonumber \\
    &-  \frac{1}{3V} \int d\br d\br' d\br'' [\mathcal{W}_{nn}]_{n^b(\br)} |\br-\br'| w(|\br-\br''|) w'(|\br-\br'|) c(\br)c(\br')c(\br'') \,\ \times \nonumber \\
    & \phantom{-  \frac{1}{3V} \int d\br d\br' d\br'' \,\ } \times \left[g(\br''|\br,\br')-g(\br'|\br)g(\br''|\br) \right] \label{Pexfields}
\end{align}
%\end{widetext}
%
where $g(\br''|\br,\br')$ is the triplet distribution function \cite{Hansen2006}. Equation \eqref{Pexfields} reveals that, at this level of approximation, three-body correlations also contribute to the evaluation of the system pressure. For simple fluids at moderate densities and without strong directionality, these correlations can be reasonably estimated using the Kirkwood superposition approximation (KSA) \cite{Kirkwood1952},
\begin{align}
    g(\br''|\br,\br') \simeq g(\br'|\br)g(\br''|\br)g(\br''|\br') \label{KSA}
\end{align}
Thus, considering a homogeneous and isotropic system, and introducing the KSA into Eq. \eqref{Pexfields}, we obtain,
%
%\begin{widetext}
\begin{align}
    P^{\text{ex}} =& - \frac{1}{3} [\mathcal{W}_n]_{n^b} \, \occ^2 \int d\br \, r \,w'(r) g(r) \,\ + \nonumber \\
    & - \frac{1}{3} [\mathcal{W}_{nn}]_{n^b} \, \occ^2 \int d\br \, r \, w(r) w'(r) g(r) \,\ + \nonumber \\
    &-  \frac{1}{3} [\mathcal{W}_{nn}]_{n^b} \, \occ^3 \int d\br d\br' \, r \, w(r') w'(r) g(r ) g(r ')\left[g(|\br ' - \br|)-1 \right]\label{Pexhomoiso}
\end{align}
%\end{widetext}
%
Equation \eqref{Pexhomoiso} relates the macroscopic pressure $P = k_BT \, \occ + P^{\text{ex}}$ to the mesoscopic parameters of the LTh model, contained in the terms $\mathcal{W}_n$ and $\mathcal{W}_{nn}$. The excess pressure is affected by the local structure of the system through the radial distribution function, which enters the expression both explicitly as well as implicitly through $n^b$.

\subsubsection{Energy EoS}

The system internal energy $U$ can be evaluated as the sum of a kinetic and a potential contribution, as \cite{ChandlerBook},
\begin{align}
    U = \Big\langle \sum_i \frac{p_i^2}{2m} \Big\rangle + \Big\langle \sum_i u(s_i,\hat{n}_i)\Big\rangle
\end{align}
The kinetic term returns the well known ideal gas contribution to the energy for a monatomic gas of mesoparticles, i.e., $(3/2)Nk_BT$. The particle internal energy term, on the other hand, needs further manipulation. From Eq. \eqref{us}, we have,
\begin{align}
    u(s_i,\hat{n}_i) = \mathcal{V}(\hat{n}_i) + C_V\Theta e^{s_i/C_V}[\psi(\hat{n}_i)]^{k_B/C_V}
\end{align}
where $\mathcal{V}(\hat{n}_i)$ and $\psi(\hat{n}_i)$ have been defined in Eqs. \eqref{pot} and \eqref{defpsi} respectively. Thus,
\begin{align}
     \Big\langle \sum_i u(s_i,\hat{n}_i)\Big\rangle = \sum_i \int & d\br^N ds^N P_{\text{eq}}(\br^N,s^N) \Big[  \mathcal{V}(\hat{n}_i) \,\ + \nonumber \\ &+ C_V\Theta e^{s_i/C_V}[\psi(\hat{n}_i)]^{k_B/C_V} \Big] \label{avgu}
\end{align}
The second term in this integral can be directly evaluated, yielding (see Appendix \ref{App-U})
\begin{align}
    \sum_i\int d\br^N ds^N P_{\text{eq}}(\br^N,s^N) C_V\Theta e^{\frac{s_i}{C_V}}[\psi(\hat{n}_i)]^{\frac{k_B}{C_V}} = NC_VT \label{psiterm}
\end{align}
By contrast, the first contribution in Eq. \eqref{avgu} requires an expansion of $\mathcal{V}(\hat{n}_i)$ analogous to Eq. \eqref{Wexpansion},
\begin{align}
    \mathcal{V}(\hat{n}_i) \simeq & \,\ [\mathcal{V}]_{n^b(\br_i)} - [\mathcal{V}_n]_{n^b(\br_i)} n^b(\br_i) \,\ + \nonumber \\
    &+ [\mathcal{V}_n]_{n^b(\br_i)} \sum_{j \neq i} w(|\br_i - \br_j|) \label{calVexpansion}
\end{align}
Inserting this approximation into Eq. \eqref{avgu} and considering a homogeneous system, one finally arrives at (see Appendix \ref{App-U}),
\begin{align}
    \sum_i \int d\br^N ds^N P_{\text{eq}}(\br^N,s^N) \mathcal{V}(\hat{n}_i) = N\mathcal{V}(n) \label{calVterm}
\end{align}
where $n$ is the average corrected local density. The EoS relating the internal energy of the system to the LTh model parameters thus reads,
\begin{align}
    U = \frac{3}{2}Nk_BT + N\mathcal{V}(n) + NC_V T \label{EoSU1}
\end{align}
or equivalently,
\begin{align}
    \frac{U}{N} =& \,\ \frac{3}{2}k_BT + \mathcal{V}(n) + C_VT \label{EoSU2}
\end{align}

\subsubsection{Parametrization of the LTh model} \label{Sub-LThpara}

The EoS Eq. \eqref{virial}, with Eq. \eqref{Pexhomoiso}, and Eq. \eqref{EoSU2} relate the macroscopic pressure and energy to the parameters of the LTh mesoscopic model. Although the mesoscopic parameters keep a close analogy with macroscopic properties, one has to keep in mind that the fluctuations include additional contributions that make, e.g., that the mesoscopic isothermal compressibility is different from the macroscopic one. Hence, to construct a truly quantitative model for macroscopic properties, the appropriate mesoscopic parameters need to be determined. Our procedure uses the theoretically derived EoS to bridge the gap between these analogous quantities. However, due to the local structure introduced by the interparticle interactions, these EoS are not directly invertible, and an iterative procedure has to be used to achieve a precise determination of the LTh parameters. Therefore, as the exact structure of the system (i.e., the $g(r)$) is not known {\em a priori}, we introduce several approximations to obtain an initial set of values for such parameters. If we consider, as a first assumption, that the fluctuations in the local density are negligible, $\delta\hat{n}_i^b = \hat{n}_i^b - n^b(\br_i) \simeq 0$, and that there are no local inhomogeneities, i.e., $g(r) \simeq 1$ everywhere (corresponding to the perspective of taking $\Rc \to \infty$), then the pressure EoS Eq. \eqref{virial} simplifies to \cite{Colella2025},
\begin{align}
    P = \occ\,k_BT + \langle\pi_i\rangle \label{virialapprox}
\end{align}
where $\langle \pi_i\rangle = \pi(T,\occ)$ under the chosen assumptions. Setting the reference mesoscopic density $n_{00} = \occ$ and temperature $\theta_0 = T$, by virtue of Eq. \eqref{simplex} we thus get a first estimation of the reference particle pressure $\pi_{00}$, namely,
\begin{align}
    \pi_{00} = P -\occ\,k_BT \label{pi00init}
\end{align}
The macroscopic isothermal compressibility, $\overline{\kappa}_T$, is also readily obtained from Eq. \eqref{virialapprox},
\begin{align}
    \frac{1}{\overline{\kappa}_T} \equiv \occ \left. \frac{\partial P}{\partial \occ} \right|_T = \occ \, k_BT + \frac{1}{\kappa_T} \label{kappamacro}
\end{align}
where use of the definition of $\kappa_T$, Eq. \eqref{k}, has been made. Equation \eqref{kappamacro} allows us to determine the mesoscopic isothermal compressibility as
\begin{align}
    \kappa_T = \frac{\overline{\kappa}_T}{1-\overline{\kappa}_T \occ \, k_BT} \label{kappainit}
\end{align}
Differentiation of Eq. \eqref{virialapprox} with respect to $T$ and $\occ$ at constant $P$ yields,
\begin{align}
    0 = \bigg(\occ\,k_B + \left.\frac{\partial \pi}{\partial T}\right|_{\occ} \bigg)dT + \bigg(k_BT + \left. \frac{\partial \pi}{\partial \occ}\right|_T \bigg)d\occ
\end{align}
Hence, the macroscopic thermal expansion coefficient, $\overline{\alpha}$, reads,
\begin{align}
    \overline{\alpha} \equiv - \frac{1}{\occ} \left. \frac{\partial \occ}{\partial T} \right|_P = \frac{\occ \, k_B + \alpha/\kappa_T}{\occ \, k_B T + 1/\kappa_T} \label{alphamacro}
\end{align}
where the definitions Eqs. \eqref{k} and \eqref{alpha} have been used. The mesoscopic thermal expansion coefficient is thus set as,
\begin{align}
    \alpha = \overline{\alpha} \, \big(1+ \kappa_T \occ \, k_BT \big) - \kappa_T \occ \, k_B \label{alphainit}
\end{align}
with $\kappa_T$ given in Eq. \eqref{kappainit}. Finally, the macroscopic heat capacity at constant volume, $\overline{C}_V$, is obtained from the energy EoS Eq. \eqref{EoSU1},
\begin{align}
    \overline{C}_V \equiv \left. \frac{\partial U}{\partial T} \right|_V =& \,\ \frac{3}{2}Nk_B + NC_V \label{CVmacro}
\end{align}
Therefore, the constant-volume heat capacity of the mesoparticle is,
\begin{align}
    C_V = \frac{\overline{C}_V}{N} - \frac{3}{2}k_B \label{CVinit}
\end{align}
Equations \eqref{pi00init}, \eqref{kappainit}, \eqref{alphainit}, and \eqref{CVinit} provide an estimation of the LTh model parameters, which can be used as an initial input for numerical simulations or to predict the $g(r)$ via the HNC approximation Eq. \eqref{HNC}. In either case, {\em a posteriori} knowledge of the system structure makes it possible to refine the mesoscopic parameters via Eqs. \eqref{Pexhomoiso} and \eqref{EoSU2}, with the aim of obtaining more accurate predictions of the macroscopic thermodynamic quantities. This iterative tuning procedure is analyzed in more detail in Section \ref{Sec-Results}.

\section{Results and discussion} \label{Sec-Results}

In this section, we present the results of equilibrium GenDPDE simulations aiming at replicating the physical behavior of argon. We begin by considering liquid conditions, specifying the selected reference state point for our tests and the associated LTh model parameters following the protocol described above. Then, we analyze the outcome of our simulations in terms of both thermodynamic quantities and local structure of the system, assessing the impact of different values of the cutoff radius $\Rc$ on the results. We also compare the system pressure $P$ and internal energy $U$ obtained numerically with the predictions from the EoS Eqs. \eqref{virial} (with Eq. \eqref{Pexhomoiso}) and \eqref{EoSU2}, respectively. Moreover, we discuss the accuracy of the HNC approach Eq. \eqref{HNC} in predicting the $g(r)$ as compared to our GenDPDE results. Finally, we investigate the range of validity of the LTh model by simulating the system at different thermodynamic conditions in the vicinity of the reference state point. We extend this latter analysis to consider also supercritical conditions, with the aim of comprehensively assessing the applicability of the LTh model to the description of condensed phases.

\subsection{Computational details} \label{Sub-CompDets}

GenDPDE simulations were performed using an OpenMP parallelized Fortran code, which was implemented by our group. Numerical integration of the EoM for the update of the particle positions, Eq. \eqref{EoMr}, and momenta, Eq. \eqref{EoMp}, was performed using a Velocity-Verlet scheme, while the EoM for the internal energy, Eq. \eqref{EoMu}, was integrated with an Euler scheme.

All tests were conducted at constant total energy (microcanonical ensemble), considering reduced units, determined via the procedure described below. The selected reference state for argon is characterized by a temperature $T_0 = 125.7$ K, mass density $\tilde{\rho}_0 = 1419.7$ kg/m$^3$, and pressure $P_0 = 85.31$ MPa \cite{NIST2024}. These conditions correspond to liquid phase, with pressure and density significantly above the respective critical values, $P_c = 4.86$ MPa and $\tilde{\rho}_c = 535.6$ kg/m$^3$, while the considered temperature is below but not far from the critical value $T_c = 150.7$ K. Hence, argon under these thermodynamic conditions represents a somewhat "hot" liquid, whose behavior should be well described by our simplified LTh model in a wider range of density excursions as compared to colder liquids, which are particularly sensitive to density variations due to their lower compressibility. The accuracy and validity of the LTh model are discussed in detail below. With these reference values, the corresponding number density of the system is,
\begin{align}
    \rho_0 = \tilde{\rho}_\text{ref} \frac{N_\text{A}}{\mathcal{M}_\text{w}} \label{rhoref}
\end{align}
with $N_\text{A}$ the Avogadro number and $\mathcal{M}_\text{w} = 4\times 10^{-2}$ kg/mol the molecular weight of the substance. The selected degree of CG in this analysis is $\phi = 5$, indicating that each mesoparticle contains $5$ physical constituents. This choice is intended to emphasize the fluctuating nature of the mesoscopic system, which would otherwise be blurred out by larger degrees of CG. The mesoscopic reference number density is thus equal to,
\begin{align}
    \overline{c}_0 = \frac{\rho_0}{\phi} \label{cref}
\end{align}
Given that each mesoparticle has a fixed mass $m$, in view of Eq. \eqref{CVinit}, the mesoparticle heat capacity at constant volume can be determined from the specific macroscopic heat capacity of argon, $\tilde{C}_{V0} = 5.20\times10^2$ J/(kg $\cdot$ K), as,
\begin{align}
    C_{V0} = \tilde{C}_{V0} \, m \label{CVpart}
\end{align}
Moreover, at the selected state point, we have that the isothermal compressibility is $\overline{\kappa}_{T0} = 1.49\times10^{-9}$ Pa\(^{-1}\), the thermal expansion coefficient is $\overline{\alpha}_0 = 2.64\times10^{-3}$ K\(^{-1}\), the kinematic viscosity is $\nu = 1.69 \times 10^{-7}$ m$^2$/s and the thermal conductivity is $\lambda = 0.139$ W/(m $\cdot$ K). The latter two properties are related to their mesoscopic counterparts, the friction coefficient $\gamma$ and the thermal conductivity $\kappa$, by theoretical estimations \cite{Marsh1997,Avalos1999,Soleymani2020},
\begin{align}
    \nu =& \,\ \frac{45 m k_B T_0}{4 \pi \gamma \Rc^3} +  \frac{2 \pi \gamma \Rc^5 \occ_0^2}{1575} \label{nupredict} \\
    \lambda =& \,\ \frac{45 C_V k_B T_0}{2 \pi \gamma \Rc^3} +  \frac{2 \pi \kappa \Rc^5 \occ_0^2}{315 T_0^2} \label{lambdapredict}
\end{align}
These expressions were calculated for ideal systems with no interparticle potential interactions. Therefore, their use to set the dissipative coefficients may introduce significant deviations in the simulated transport coefficients, with respect to the predicted ones. However, the analysis of the dynamic properties lies outside the scope of the present work. 

Let us now introduce the reference scales for the length, energy, mass and time, which will allow us to convert all the physical quantities of interest for our simulations into reduced units. The scale of length is chosen to be
\begin{align}
    L_\text{ref} = \left( \frac{1}{\overline{c}_0} \right)^\frac{1}{3} \label{Lreference}
\end{align}
which automatically sets the dimensionless number density of the mesoscopic system to $\overline{c}_0^* = 1$, by construction. The unit of energy is
\begin{align}
    u_\text{ref} = \frac{L_\text{ref}^3}{\overline{\kappa}_{T0}} \label{ureference}
\end{align}
This selection parallels the choice of Ref. \cite{Colella2025} to determine the energy scale from potential interactions, in analogy to what is typically done in MD approaches. However, notice that in DPD models the usual choice is $k_BT$ as the scale of energy. Finally, the unit of mass is selected as,
\begin{align}
    m_\text{ref} = \phi \frac{\mathcal{M}_\text{w}}{N_\text{A}} \label{mreference}
\end{align}
that corresponds to the mass of a mesoparticle as obtained from the mass of its physical constituents. From these reference units, the time scale can be derived as,
\begin{align}
    t_\text{ref} = \sqrt{\frac{m_\text{ref} L_\text{ref}^2}{u_\text{ref}}} \label{treference}
\end{align}
Thus, in reduced units, $\Rc^* \equiv \Rc/L_\text{ref}$, $n_i^* \equiv n_iL_\text{ref}^3$, $u_i^* \equiv u_i/u_\text{ref}$, $\theta_i^* \equiv k_B\theta_i/u_\text{ref}$,  $\pi_i^* \equiv \pi_i L_\text{ref}^3/u_\text{ref} \equiv \pi_i \overline{\kappa}_{T0}$, $\kappa_T^* \equiv \kappa_T/\overline{\kappa}_{T0}$, $\alpha^* \equiv \alpha \, u_\text{ref}/k_B$, $C_{V,i}^* \equiv C_{V,i}/k_B$, $\nu^* \equiv \nu \, t_\text{ref}/L_\text{ref}^2$, $\lambda^* \equiv \lambda \, t_\text{ref} L_\text{ref}/k_B$, $m_i^* \equiv m_i/m_\text{ref}$, and $t^* \equiv t/t_\text{ref}$. Reduced units are always indicated by an asterisk $(^*)$ in what follows.

Having defined the reference scales, and following the procedure of Section \ref{Sub-LThpara}, the dimensionless state-point quantities and LTh parameters to be used in our simulations of liquid argon can be readily calculated. These values are summarized in Table \ref{tab-refvals} below.
\begin{table}[H]
    \centering
    \caption{Dimensionless state-point values and LTh parameters for GenDPDE simulations of liquid argon at $T_0 = 125.7$ K, $\tilde{\rho}_0 = 1419.7$ kg/m$^3$, and $P_0 = 85.31$ MPa.}
    %\vspace{2mm}
    \begin{tabularx}{\linewidth}{Y|Y|Y|Y|Y|Y|Y|Y|Y}
    \toprule
        $T_0^*$ & $P_0^*$ & $\occ_0^*$ & $C_{V0}^*$ & ${\kappa}_{T0}^*$ & ${\alpha}_{0}^*$ & $\theta_{0}^*$ & $\pi_{00}^*$ & $n_{00}^*$\\
    \midrule
        $0.0111$ & $0.127$ & $1.0$ & $12.51$ & $1.01$ & $29.35$ & $0.0111$ & $0.1161$ & $1.0$ \\
    \bottomrule
    \end{tabularx}
    \label{tab-refvals}
\end{table}

According to Eqs. \eqref{nupredict} and \eqref{lambdapredict}, the mesoscopic dynamic coefficients, $\gamma^*$ and $\kappa^*$, depend on the magnitude of the selected cutoff distance. In Table \ref{tab-gammakappa}, their values are reported for the different $\Rc^*$ considered in our tests.
\begin{table}[H]
    \centering
    \caption{Dimensionless mesoscopic friction coefficient $\gamma^*$ and thermal conductivity $\kappa^*$ as functions of the cutoff distance $\Rc^*$.}
    %\vspace{2mm}
    \begin{tabularx}{\linewidth}{Y|Y|Y}
    \toprule
        $\Rc^*$ & $\gamma^*$ & $\kappa^*$ \\
    \midrule
        $1.3365$ & $23.42$ & $0.0079$ \\
        $1.6839$ & $7.37$ & $0.0025$ \\
        $2.1564$ & $2.14$ & $0.00072$ \\
    \bottomrule
    \end{tabularx}
    \label{tab-gammakappa}
\end{table}

Equilibrium simulations were performed considering a cubic box of $N=27000$ mesoparticles, with periodic boundary conditions (PBC). All tests considered an initial equilibration period of $1000$ time units, in which velocity rescaling was applied to make the system reach the target nominal temperature, followed by a production run of $2000$ time units. Energy conservation was ensured in all cases by properly adjusting the size of the time step and, therefore, the number of iterations required to complete the simulations. These values are reported as a function of the cutoff distance $\Rc^*$ in Table \ref{tab-timestep}.
\begin{table}[H]
    \centering
    \caption{Time step, $\delta t^*$, number of equilibration iterations, $N_{\text{eq}}$, number of production run iterations, $N_{\text{pr}}$, and total number of iterations, $N_{\text{tot}}$, as functions of the cutoff distance $\Rc^*$.}
    %\vspace{2mm}
    \begin{tabularx}{\linewidth}{Y|Y|Y|Y|Y}
    \toprule
        $\Rc^*$ & $\delta t^*$ & $N_{\text{eq}}$ & $N_{\text{pr}}$ & $N_{\text{tot}}$ \\
    \midrule
        $1.3365$ & $5\times10^{-4}$ & $2\times10^{6}$ & $4\times10^{6}$ & $6\times10^{6}$ \\
        $1.6839$ & $1\times10^{-3}$ & $1\times10^{6}$ & $2\times10^{6}$ & $3\times10^{6}$ \\
        $2.1564$ & $1\times10^{-3}$ & $1\times10^{6}$ & $2\times10^{6}$ & $3\times10^{6}$ \\
    \bottomrule
    \end{tabularx}
    \label{tab-timestep}
\end{table}

Finally, the scaling factor $f_{\text{cut}}$ involved in the evaluation of the corrected local density Eq. \eqref{newvol} was adjusted according to the value of $\Rc^*$ as shown in Table \ref{tab-fcut}, following Ref. \cite{Colella2025}.
\begin{table}[H]
    \centering
    \caption{Scaling factor $f_{\text{cut}}$ as a function of the cutoff distance $\Rc^*$.}
    %\vspace{2mm}
    \begin{tabularx}{\linewidth}{Y|Y}
    \toprule
        $\Rc^*$ & $f_{\text{cut}}$ \\
    \midrule
        $1.3365$ & $1.41$ \\
        $1.6839$ & $1.35$ \\
        $2.1564$ & $1.33$ \\
    \bottomrule
    \end{tabularx}
    \label{tab-fcut}
\end{table}

\subsection{Results}

The average kinetic temperature $T^*$, corrected number density $n^*$, pressure $P^*$, and internal energy per particle $U^*/N$, obtained from GenDPDE tests with the input parameters specified in Tables \ref{tab-refvals} to \ref{tab-fcut}, are reported in Table \ref{tab-thermo1} below. The system pressure has been calculated as an average over instantaneous values, evaluated in the code using the Virial formula
\begin{align}
    P \equiv \frac{1}{3V} \biggl \langle \sum_i \frac{p_i^2}{m_i} + \sum_i \sum_{j<i} \br_{ij} \cdot \bff_{ij}^C \biggr \rangle \label{SimVirial}
\end{align}
It can be observed that, in all cases, the numerical temperature $T^*$ is satisfactorily close to the expected nominal value $T_0^*=0.0111$, while the numerical density $n^*$ slightly overestimates the nominal $\occ_0^*=1.0$, but the accuracy improves as $\Rc^*$ increases. In all tests, the pressure $P^*$ estimated using the LTh model is also close to the nominal $P_0^*=0.127$, although it remains slightly underestimated. This indicates that a refined calibration of the initial set of model parameters, in particular of the reference pressure $\pi_{00}^*$, can further improve the quantitative accuracy of GenDPDE simulations. Such a fine-tuning procedure is thoroughly discussed in the subsequent section. Finally, notice that the nominal value for the internal energy $U^*$ for argon, as given in the NIST Webbook, is not provided here for comparison, since it is calculated with respect to a zero-energy state that cannot be attained within our model. However, as energy differences are actually comparable, in Section \ref{RoV} we evaluate the energy variations for conditions around our reference state point, comparing the experimental data with our numerical results. Instead, in this section, we compare the values of $U^*/N$ obtained from simulations at the reference state with the theoretical predictions based on the same model and energy origin, to test the predictive capacity of the EoS Eq. \eqref{EoSU2}.
\begin{table}[H]
    \centering
    \caption{Average temperature $T^*$, number density $n^*$, pressure $P^*$, and internal energy per particle $U^*/N$ from GenDPDE simulations as functions of the cutoff distance $\Rc^*$. }
    %\vspace{2mm}
    %{\fontsize{7.6}{11}\selectfont
    \begin{tabularx}{\linewidth}{Y|Y|Y|Y|Y}
    \toprule
        $\Rc^*$ & $T^*$ & $n^*$ & $P^*$ & $\frac{U^*}{N}$ \\
    \midrule
        $1.3365$ & $0.0112 \pm 0.0001$ & $1.007 \pm 0.001$ & $0.121 \pm 0.001$ & $-0.62 \pm 0.01$ \\
        $1.6839$ & $0.0112 \pm 0.0001$ & $1.003 \pm 0.001$ & $0.116 \pm 0.001$ & $-0.62 \pm 0.01$ \\
        $2.1564$ & $0.0111 \pm 0.0001$ & $1.001 \pm 0.001$ & $0.119 \pm 0.001$ & $-0.63 \pm 0.01$ \\
    \bottomrule
    \end{tabularx}
    %}
    \label{tab-thermo1}
\end{table}

\begin{figure}[H]
    \centering
    \subfloat{
    \includegraphics[width=0.8\linewidth]{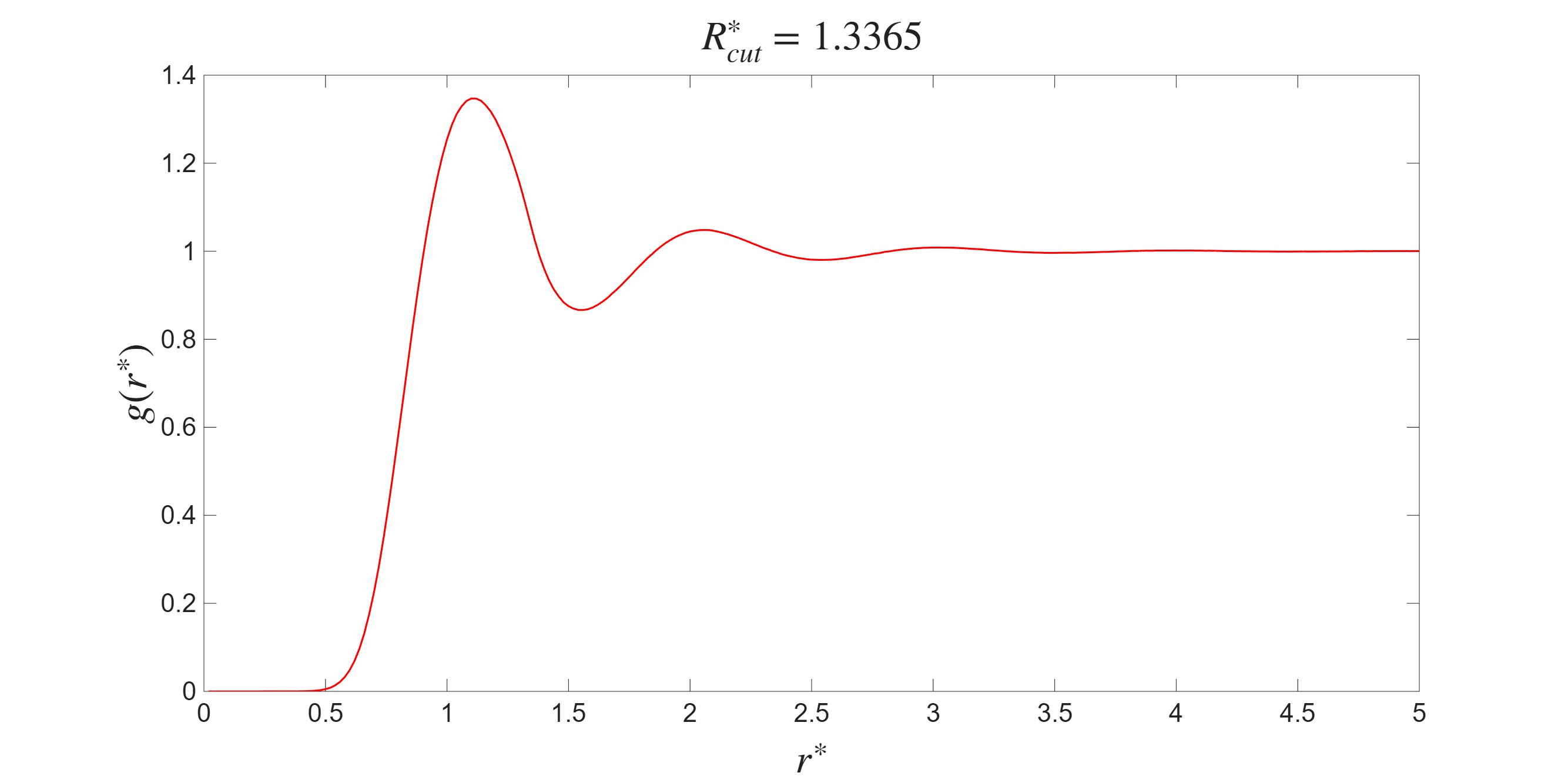}
    } \hfill
    \subfloat{
    \includegraphics[width=0.8\linewidth]{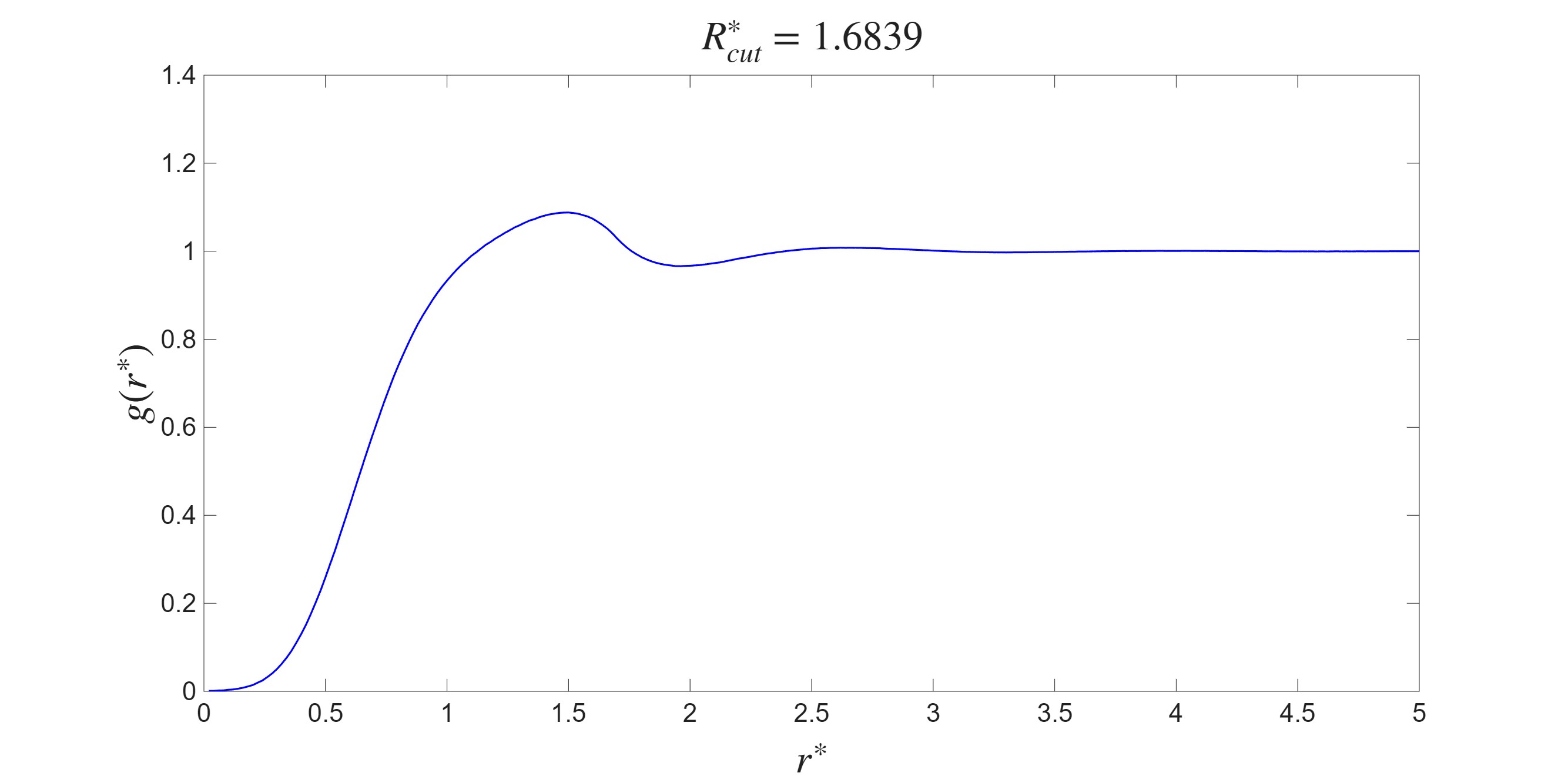}
    } \hfill
    \subfloat{
    \includegraphics[width=0.8\linewidth]{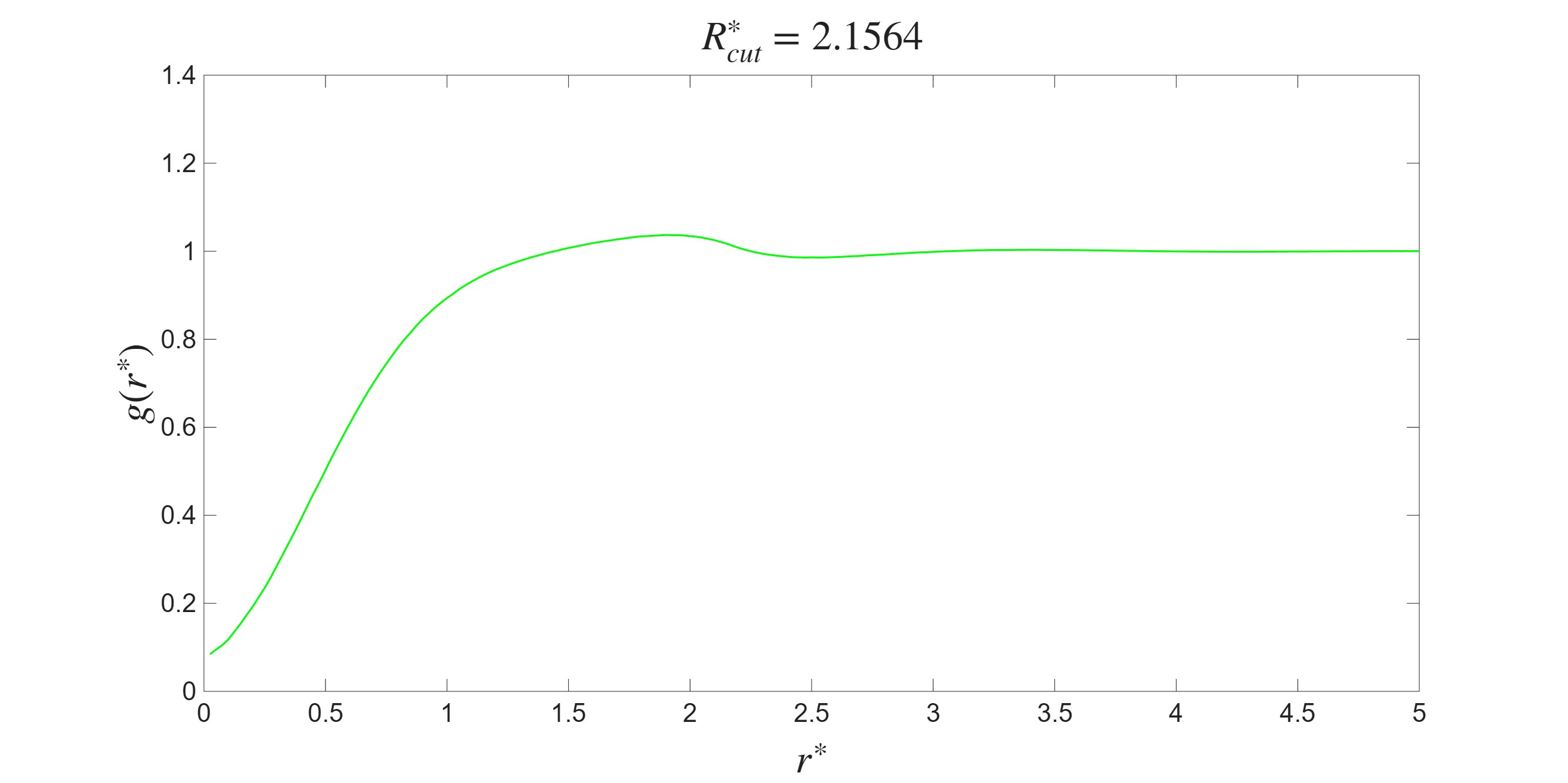}
    }
    \caption{Argon radial distribution functions for $\Rc^* = 1.3365$ (top), $\Rc^* = 1.6839$ (middle), $\Rc^* = 2.1564$ (bottom) from GenDPDE simulations.}
    \label{fig-RDF1}
\end{figure}

The radial distribution functions for all GenDPDE tests are shown in Fig. \ref{fig-RDF1}. As anticipated, knowledge of the $g(r^*)$ allows us to estimate the system pressure $P^*$ and internal energy per particle $U^*/N$ from the theoretical EoS, Eqs. \eqref{virial} (with \eqref{Pexhomoiso}) and \eqref{EoSU2}. The results of these predictions are reported in Table \ref{tab-EoSGenDPDE}. Here, the average density $n^*$ is also included, where this quantity is also calculated through an analytical route, using the alternative volume estimator Eq. \eqref{eq:1iv}, with $n^b$ obtained from knowledge of the $g(r^*)$ via Eq. \eqref{local2}. In all cases, we observe that the average density is satisfactorily close to its nominal value $\occ_0^*=1.0$, with errors $\le 1\%$. For $\Rc=1.6839$ and $\Rc=2.1564$, we also find that the predicted pressure is less than $3.5\%$ away from the nominal $P_0^*=0.127$, and that the internal energy per particle agrees well the respective simulation values reported in Table \ref{tab-thermo1}, with errors below $2\%$. For $\Rc^*=1.3365$, on the other hand, we observe that the predicted pressure overestimates the nominal value by $\sim 8\%$. The reason behind this reduced accuracy is to be found in the shape of the $g(r^*)$, which for $\Rc^*=1.3365$ is characterized by more pronounced peaks and a substantial {\em correlation hole} surrounding the central particle, with a width $r^* \simeq 0.5$. This is indicative of the fact that, as $\Rc^*$ decreases, correlations between mesoparticles become stronger. Moreover, in this limit, the magnitude of the fluctuations in the instantaneous density also increases, yielding stronger repulsive forces. Thus, these results reveal that, for small $\Rc^*$, the approximations underlying Eq. \eqref{Pexhomoiso} start to fall outside their range of validity. In such situations, with the aim of improving the accuracy of the results, higher-order expansions in the density for the potential $\mathcal{W}$ may be needed, and the KSA should be enhanced by, e.g., use of a correction factor that better accounts for three-body contributions \cite{Abe1959}.
\begin{table}[H]
    \centering
    \caption{Average number density $n^*$, and average pressure $P^*$ and internal energy per particle $U^*/N$ from EoS predictions Eqs. \eqref{virial}-\eqref{Pexhomoiso} and \eqref{EoSU2}, calculated using the $g(r^*)$ from GenDPDE simulations, as functions of the cutoff distance $\Rc^*$.}
    %\vspace{2mm}
    \begin{tabularx}{\linewidth}{Y|Y|Y|Y}
    \toprule
        $\Rc^*$ & $n^*$ & $P^*$ & $\frac{U^*}{N}$ \\
    \midrule
        $1.3365$ & $0.990$ & $0.137$ & $-0.61$ \\
        $1.6839$ & $0.993$ & $0.126$ & $-0.61$ \\
        $2.1564$ & $0.995$ & $0.123$ & $-0.62$ \\
    \bottomrule
    \end{tabularx}
    \label{tab-EoSGenDPDE}
\end{table}

\subsubsection{Fine-tuning of the mesoscopic parameters} \label{Sub-FineTune}

\begin{figure}[H]
    \centering
    \includegraphics[width=0.8\linewidth]{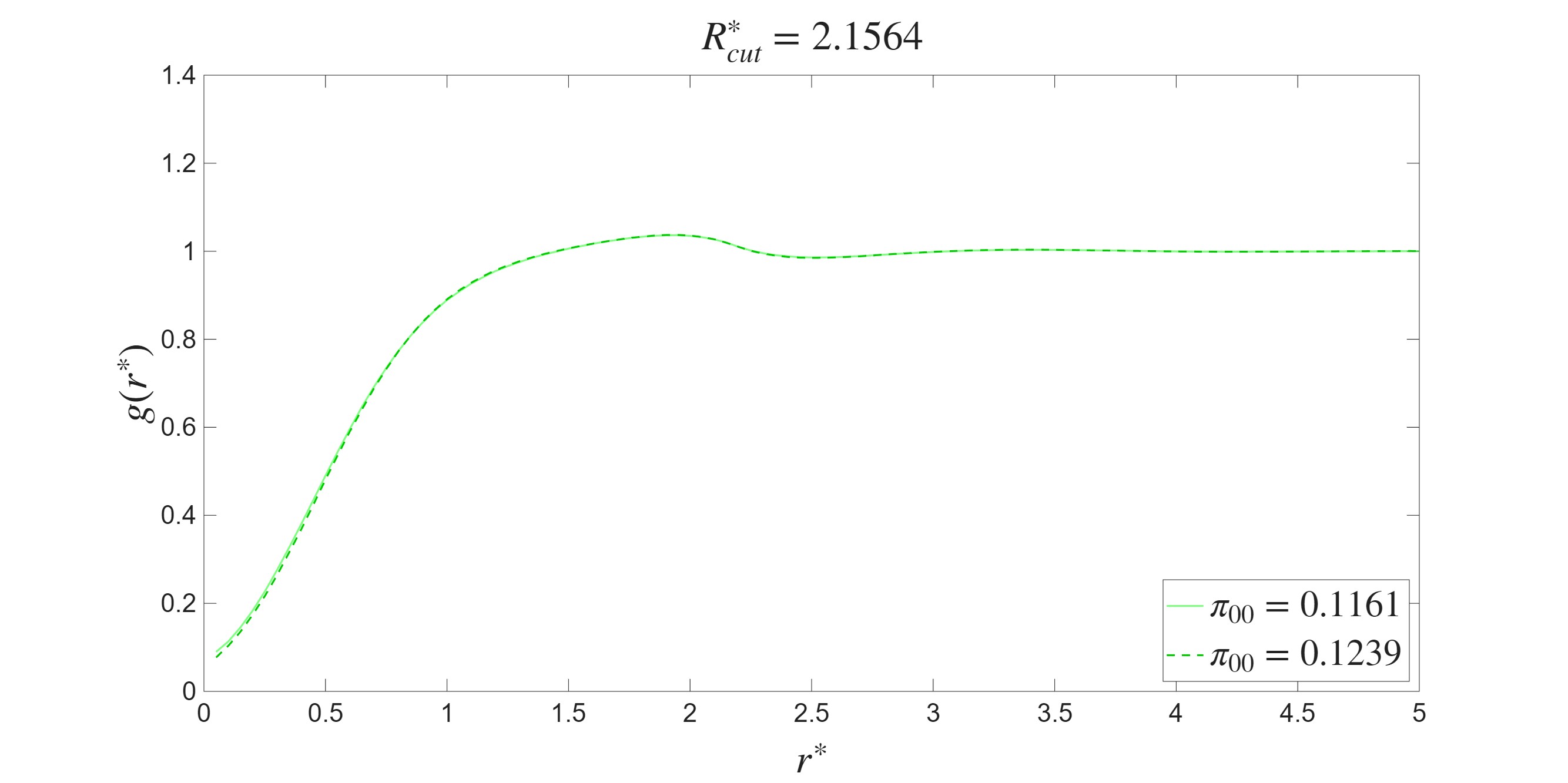}
    \caption{Argon radial distribution functions for $\Rc^* = 2.1564$ from GenDPDE tests considering an input particle pressure $\pi_{00}^*=0.1239$ (continuous line) and $\pi_{00}^*=0.1161$ (dashed line).}
    \label{fig-RDF2}
\end{figure}

The results in Table \ref{tab-thermo1} highlight the inaccuracies in measurements of the system pressure from GenDPDE simulations using an initial set of approximated LTh model parameters. Nevertheless, the discrepancy between the numerical results and the expected nominal values can be eliminated by suitably refining such parameters. More precisely, it is sufficient to fine-tune the particle reference pressure, $\pi_{00}^*$, increasing its value by a quantity $\Delta\pi^* = P_0^*-P^*$, to improve the results of GenDPDE tests. Table \ref{tab-thermo2} shows the outcome of a GenDPDE simulation considering $\Rc^* = 2.1564$ and a modified particle reference pressure $\pi_{00}^*=0.1239$. When compared with the values in Table \ref{tab-thermo1}, these results prove the beneficial effects of the fine-tuning on the evaluation of both, the system pressure and number density, which now match the expected nominal values, $P_0^*$ and $\occ_0^*$, respectively. On the other hand, the internal energy per particle is not affected by the tuning procedure. Finally, fig. \ref{fig-RDF2} shows that the fine-tuning does not alter the local structure either, as the radial distribution functions of the original and refined simulations overlap.
\begin{table}[H]
    \centering
    \caption{Average temperature $T^*$, number density $n^*$, pressure $P^*$, and internal energy per particle $U^*/N$ from a GenDPDE simulation considering $\pi_{00}^*=0.1239$.}
    %\vspace{2mm}
    %{\fontsize{7.6}{11}\selectfont
    \begin{tabularx}{\linewidth}{Y|Y|Y|Y|Y}
    \toprule
        $\Rc^*$ & $T^*$ & $n^*$ & $P^*$ & $\frac{U^*}{N}$ \\
    \midrule
        $2.1564$ & $0.0111 \pm 0.0001$ & $1.000 \pm 0.001$ & $0.127 \pm 0.001$ & $-0.63 \pm 0.01$ \\
    \bottomrule
    \end{tabularx}
    %}
    \label{tab-thermo2}
\end{table}

\subsubsection{The Hypernetted Chain approximation}
In the previous sections, we have seen that the macroscopic EoS Eqs. \eqref{virial}, \eqref{Pexhomoiso} and \eqref{EoSU2} relate the system pressure and internal energy to the parameters of the mesoscopic LTh model. We have also seen that the radial distribution function plays a non-negligible role in these relationships, even at the lowest level of approximation. However, as the exact shape of the $g(r^*)$ is not known {\em a priori}, in Section \ref{Sub-LThpara} we have introduced a few additional assumptions to obtain an initial set of parameters for our GenDPDE simulations. Then, using the outcome of these tests, we have applied a fine-tuning procedure to improve the accuracy of our GenDPDE results. At this point, one may wonder whether there is a way to obtain the exact form of the $g(r^*)$ without resorting to GenDPDE tests, and thus obtain an accurate set of LTh parameters directly from the analytical EoS. In this work, we have explored the application of the Hypernetted Chain (HNC) approximation as a tool to predict the radial distribution function in systems described by many-body potentials. This procedure was first introduced in Ref. \cite{Merabia2007}, although here we extend its use to density- and temperature-dependent potentials. Similarly to Ref. \cite{Merabia2007}, in Eq. \eqref{Wexpansion2} we have expressed the many-body potential $\mathcal{W}$ as an approximated pairwise potential, plus a one-body contribution. For a homogeneous and isotropic system, we can reformulate Eq. \eqref{Wexpansion2} in a more compact way, as,
\begin{align}
    \mathcal{W}(\hat{n}_i) \simeq u_0 + \sum_{j \neq i} u_1(|\br_i-\br_j|)  \label{defu}
\end{align}
where $u_0 \equiv \mathcal{W}(n) - [\mathcal{W}_n]_{n^b} \, n^b$ is now a constant contribution, and $u_1(|\br_i-\br_j|) \equiv [\mathcal{W}_n]_{n^b} \,w(|\br_i-\br_j|)$. In Ref. \cite{Merabia2007}, it is demonstrated that the HNC approach on the approximated pairwise potential produces accurate results for a translationally invariant and isotropic system. Therefore, here we numerically solve the HNC equation, 
\begin{align}
    \ln g(r^*_{12}) + \beta u_1(r^*_{12}) = \occ \int d\br_3 &[ h(r^*_{13}) - \ln g(r^*_{13}) - \nonumber \\
    &- \beta u_1(r^*_{13})] h(r^*_{23}) \label{HNC}
\end{align}
for the potential Eq. \eqref{defu}. Here, $h(r^*)$ is the total correlation function, defined as $h(r^*) \equiv g(r^*)-1$. Moreover, $r^*_{12}$ stands for the interparticle distance $r^*_1-r^*_2$ where the subscripts $1$ and $2$ indicate different arbitrary particles, due to the homogeneity of the system, and similarly with $3$. Finally, notice that the one-particle potential contribution $u_0$ does not appear in Eq. \eqref{HNC} as it is a constant eventually absorbed by the normalization of the distributions.

In comparison with molecular systems, here Eq. \eqref{HNC} depends on an arbitrary function $g_\text{ref}(r^*)$ that defines the values of the parameters in both the potential $u_0$ and $u_1$. As Eq. \eqref{HNC} is valid independently of the specific functional form of $g_\text{ref}(r^*)$, we choose to identify the latter with the solution for $g(r^*)$ of Eq. \eqref{HNC}, in a self-consistent manner. Furthermore, notice that, by contrast with the so-called Reference-HNC (RHNC) equation \cite{Lado1983}, here we do not intend to perform a perturbative calculation of the thermodynamic properties via thermodynamic integration, which needs an a posteriori thermodynamic consistency treatment. Instead, the reference $g_\text{ref}(r^*)$ parametrically enters our problem solution, due to its effect on the scalar coefficients that define the actual potential. The free energy is well defined for the system represented through the approximate potential of Eq. \eqref{defu}, and therefore there is no need for additional information to maintain the thermodynamic consistency, in the way it is addressed in RHNC.

\begin{figure}[H]
    \centering
    \subfloat{
    \includegraphics[width=0.8\linewidth]{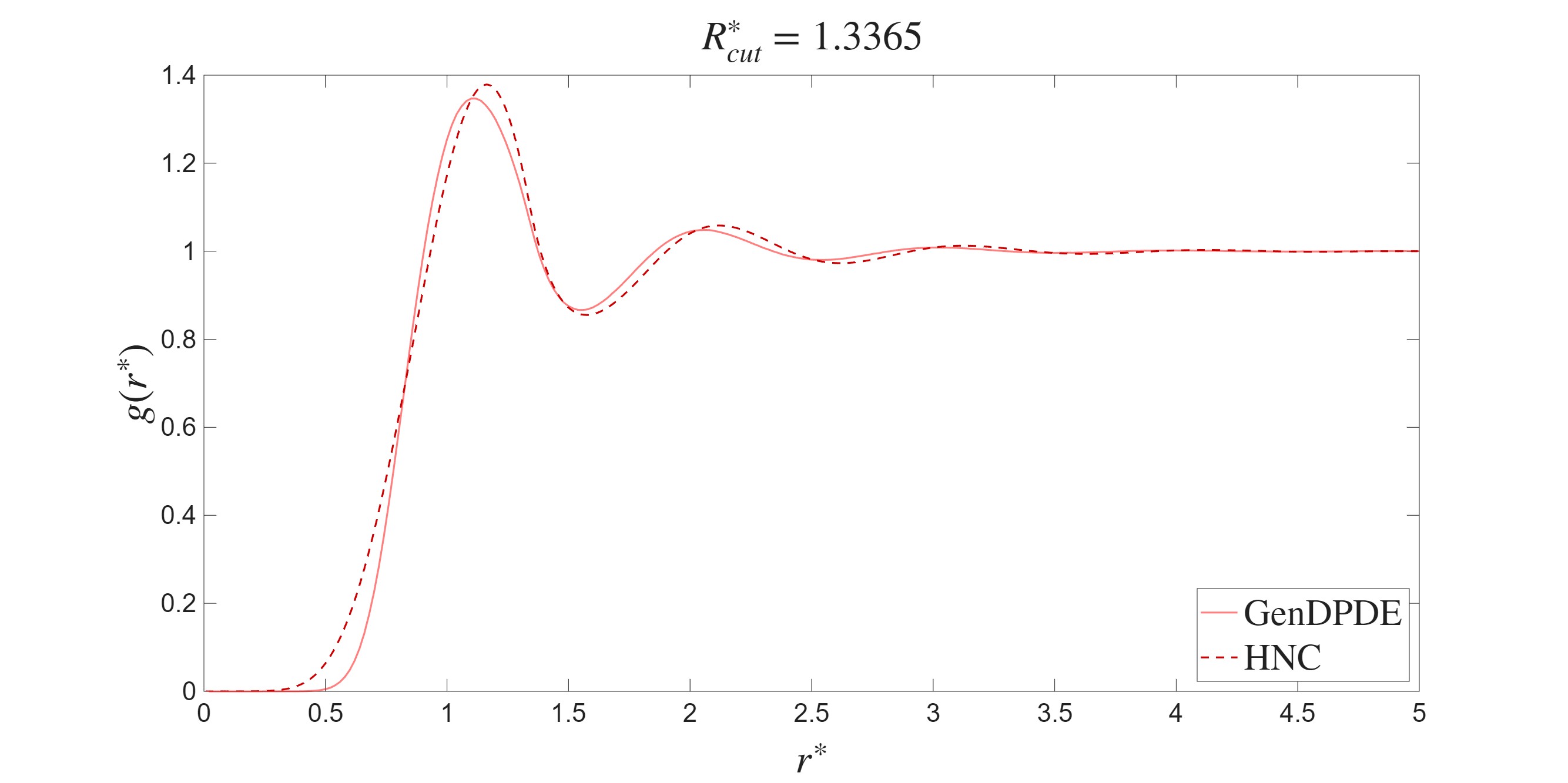}
    } \hfill
    \subfloat{
    \includegraphics[width=0.8\linewidth]{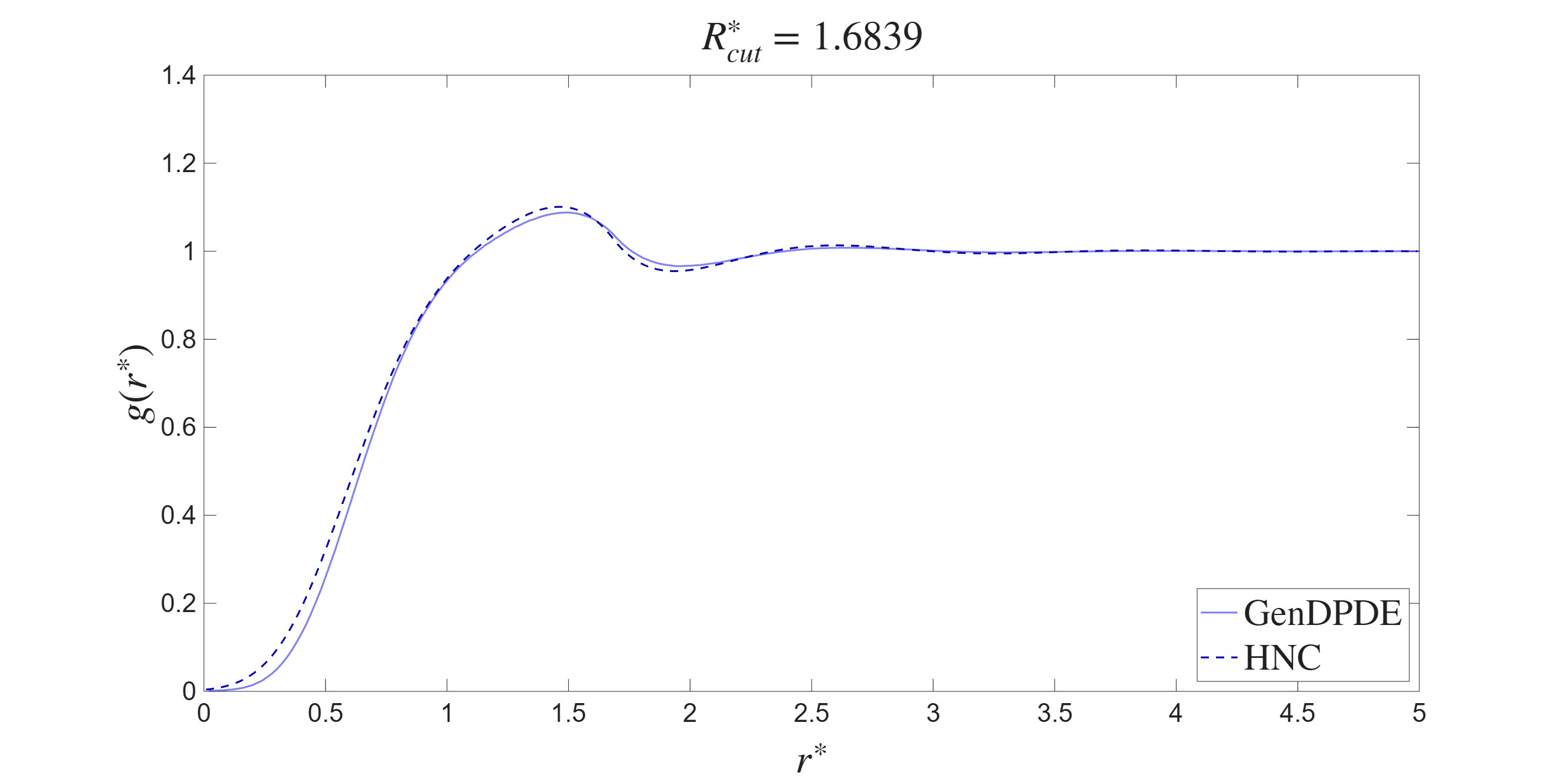}
    } \hfill
    \subfloat{
    \includegraphics[width=0.8\linewidth]{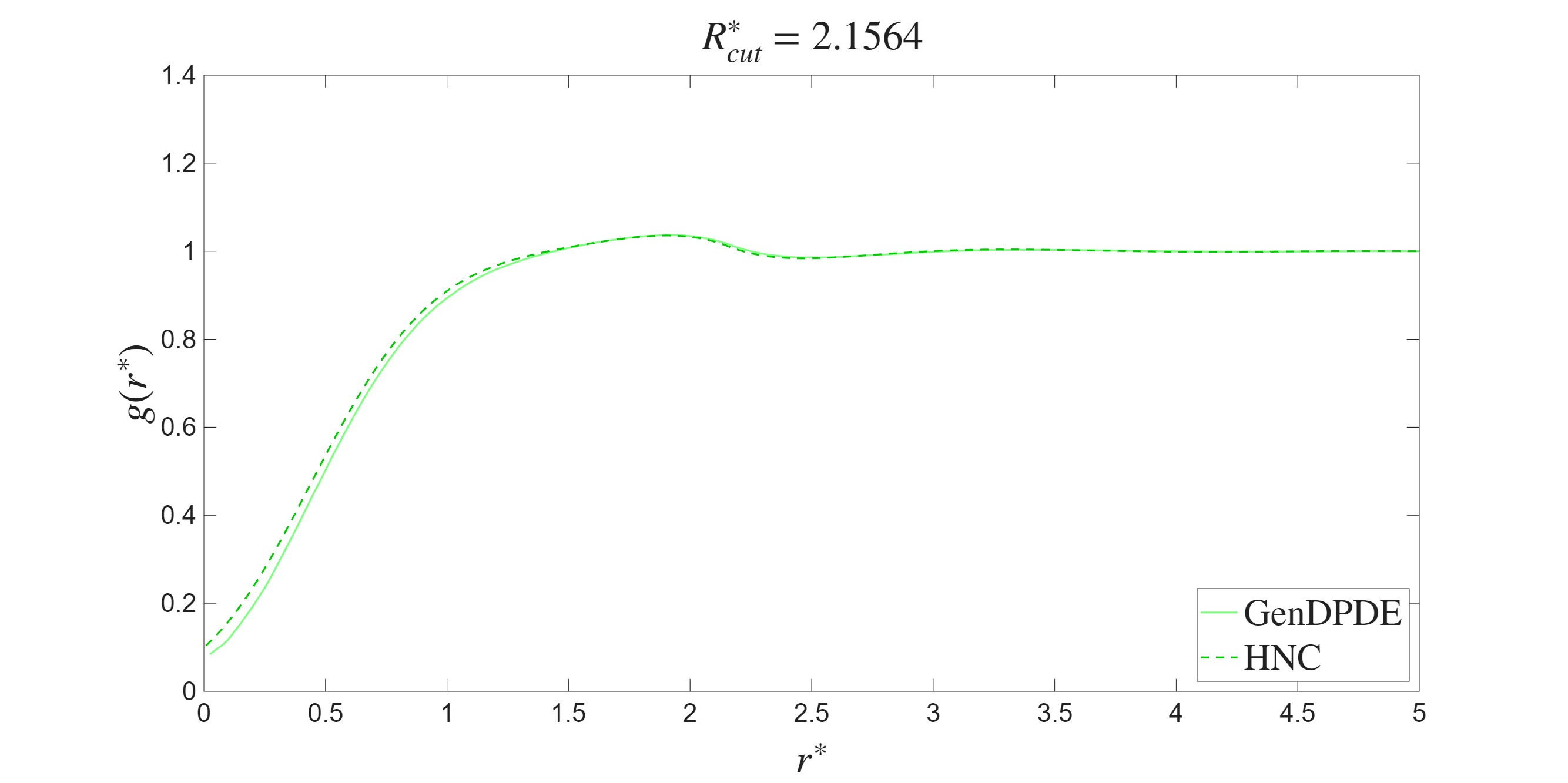}
    }
    \caption{Argon radial distribution functions for $\Rc^* = 1.3365$ (top), $\Rc^* = 1.6839$ (middle), $\Rc^* = 2.1564$ (bottom). The HNC predictions (dashed lines) Eq. \eqref{HNC} are compared against GenDPDE results (solid lines).}
    \label{fig-RDF3}
\end{figure}

In Fig. \ref{fig-RDF3}, the $g(r^*)$ obtained from our HNC predictions is compared against GenDPDE results for the different cutoff distances considered in our tests ({\em cf.} Fig. \ref{fig-RDF1}). We observe satisfactory qualitative agreement, which confirms the accuracy of the HNC approximation in predicting the system structure for many-body potentials \cite{Merabia2007}. 
Nevertheless, when used as a quantitative tool in the macroscopic EoS Eqs. \eqref{Pexhomoiso} and \eqref{EoSU2}, the HNC predictions tend to overestimate quite significantly the system pressure, as compared to the expected state-point value $P_0^*=0.127$. The nominal density, $\occ_0^*=1.0$, is also overestimated. On the other hand, the internal energy per particle is consistently underestimated with respect to the simulation results (see Table \ref{tab-thermo1}). These results highlight a quantitative limitation of the HNC approach in the present context: although it captures the system structure satisfactorily at the qualitative level, its accuracy is not sufficient to provide, on its own, the level of precision required for the direct determination of the LTh parameters entering quantitative GenDPDE simulations. For instance, according to the value of $P^*$ for $R_{\mathrm{cut}}^* = 2.1564$ in Table~\ref{tab-EoSHNC}, the corresponding estimate of the reference particle pressure $\pi_{00}^*$ would need to be adjusted by $\Delta \pi^* = -0.0088$. However, as shown in Section~\ref{Sub-FineTune}, the GenDPDE simulations indicate that $\pi_{00}^*$ must instead be increased in order to recover the target reference pressure. Therefore, within the present workflow, the HNC approximation is best viewed as a valuable structural tool, while quantitative parameter refinement is still conveniently performed through the calibration procedure described in Section~\ref{Sub-FineTune}.
%This represents an important limitation of the HNC approach, as it prevents its use for an {\em a priori} calculation of the LTh model parameters, with the aim of obtaining more accurate input values for the GenDPDE simulations. Indeed, according to, e.g., the value of $P^*$ for $\Rc^* = 2.1564$ in Table \ref{tab-EoSHNC}, the reference particle pressure $\pi_{00}^*$ needs to be adjusted by a $\Delta\pi^* = -0.0088$ in order to make up for the overestimation of the HNC prediction. Nevertheless, in Section \ref{Sub-FineTune} we clearly demonstrated that $\pi_{00}^*$ should actually be increased to correct the underestimation of $P_0^*$ from the GenDPDE test. In conclusion, unless a more precise quantitative tool for the prediction of the $g(r^*)$ is developed for density- and temperature-dependent potentials, we are obliged to apply the fine-tuning procedure described in Section \ref{Sub-FineTune} in order to improve the accuracy of our GenDPDE simulations.
%
\begin{table}[H]
    \centering
    \caption{Average number density $n^*$, and average pressure $P^*$ and internal energy per particle $U^*/N$ from EoS predictions Eqs. \eqref{virial}-\eqref{Pexhomoiso} and \eqref{EoSU2}, calculated using the $g(r^*)$ obtained from the HNC approximation Eq. \eqref{HNC}, as functions of the cutoff distance $\Rc^*$.}
    %\vspace{2mm}
    \begin{tabularx}{\linewidth}{Y|Y|Y|Y}
    \toprule
        $\Rc^*$ & $n^*$ & $P^*$ & $\frac{U^*}{N}$ \\
    \midrule
        $1.3365$ & $1.034$ & $0.169$ & $-0.70$ \\
        $1.6839$ & $1.019$ & $0.148$ & $-0.67$ \\
        $2.1564$ & $1.010$ & $0.136$ & $-0.65$ \\
    \bottomrule
    \end{tabularx}
    \label{tab-EoSHNC}
\end{table}

\subsubsection{Range of validity of the LTh model} \label{RoV}

\begin{figure}[H]
    \centering
    \subfloat{
    \includegraphics[width=0.8\linewidth]{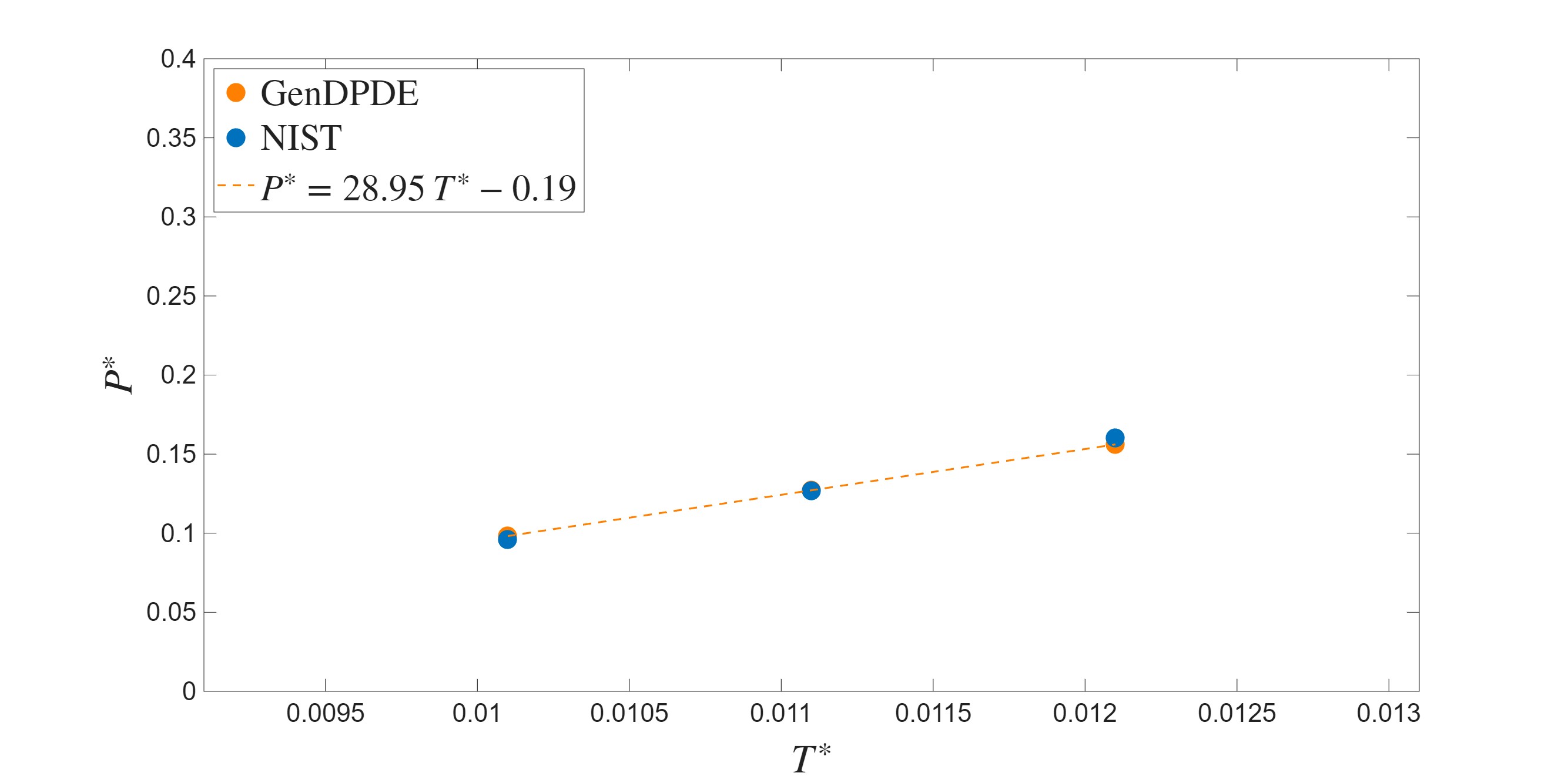}
    } \hfill
    \subfloat{
    \includegraphics[width=0.8\linewidth]{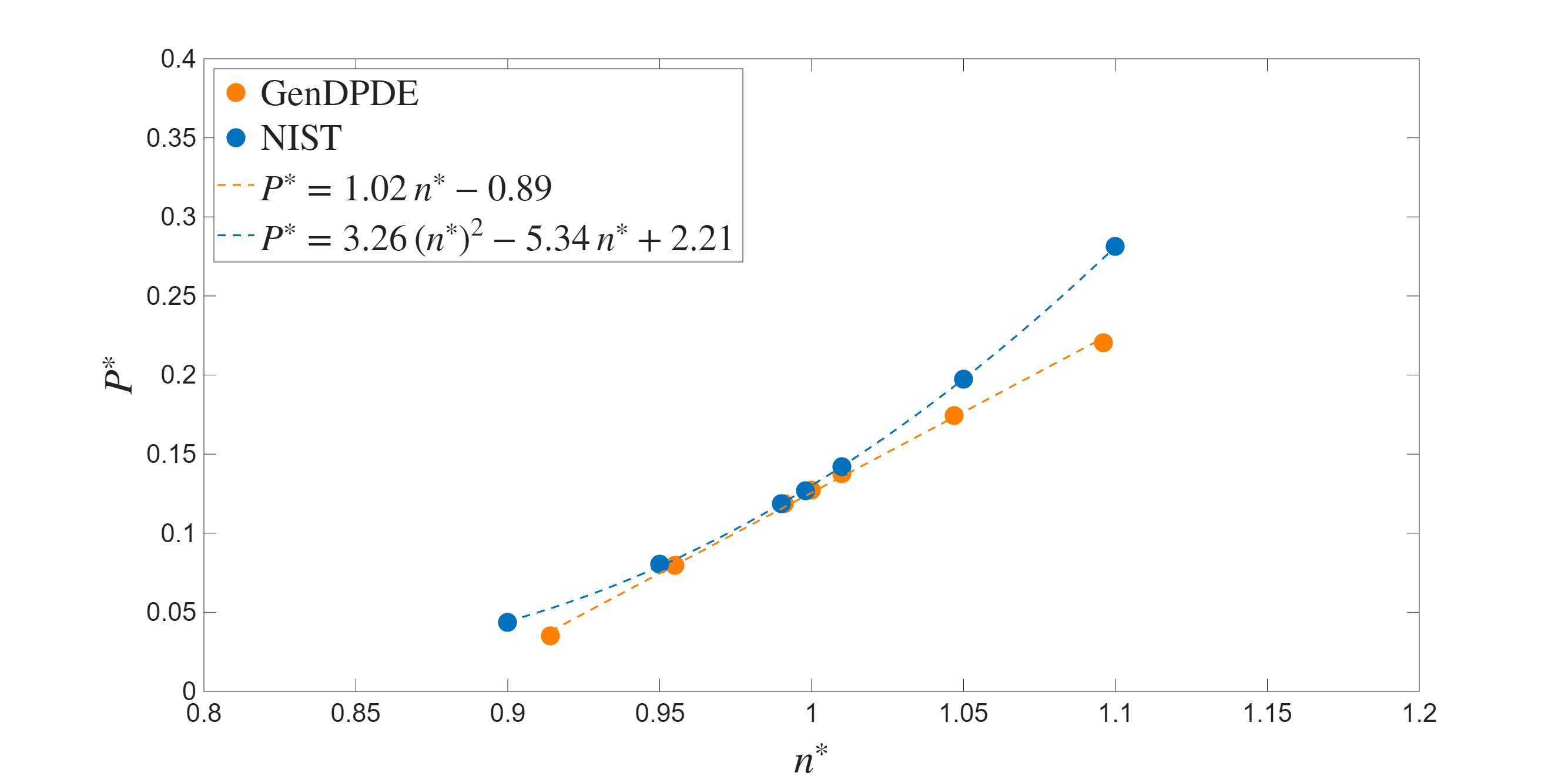}
    } \hfill
    \caption{System pressure $P^*$ as a function of the temperature $T^*$ (top) and number density $n^*$ (bottom), as obtained from GenDPDE simulations of liquid argon considering $\Rc^* = 2.1564$ and $\pi_{00}^*=0.1239$. The central point in both diagrams refers to the simulation at the reference state point ({\em cf.} Table \ref{tab-thermo2}). The numerical results are compared with data available from NIST \cite{NIST2024}.}
    \label{fig-PvsTc}
\end{figure}

As a final analysis, we examine the response of the LTh model to variations in temperature and number density around the reference state, with the aim of assessing the thermodynamic range over which the model remains quantitatively reliable. This analysis is relevant for the simulation of, e.g., non-equilibrium scenarios in which both temperature and density gradients are present. To this end, we performed a set of 8 simulations in the vicinity of the reference state, considering $\Rc^*=2.1564$ and the fine-tuned $\pi_{00}^*=0.1239$. In each of these simulations, either the nominal temperature $T_0^*$ or number density $\occ_0^*$ was changed as an input parameter, while keeping all the other quantities at their reference values ({\em cf.} Table \ref{tab-refvals}). The response of the model, in terms of measured pressure $P^*$, is reported in Fig. \ref{fig-PvsTc}, where we make a comparison between the numerical results and data available from NIST \cite{NIST2024}. We observe that, while the system responds well to temperature variations up to $\Delta T^* = \pm 10\% \, T_0^*$, it is more sensitive to density changes. More specifically, for density variations of the order of $\Delta c_0^* = \pm 1\% \, \occ_0^*$ ($\Delta n^* = \pm 1\% \, n_0^*$), we find that simulations are in excellent agreement with the experiments. However, for excursions $\Delta c_0^* = \pm 10\% \, \occ_0^*$ the difference between numerical results and NIST data becomes noticeable, with errors of about $25\%$, corresponding to large variations in pressure up to $\Delta P \simeq 40$ MPa. Actually, this behavior is to be expected in weakly compressible liquids. Indeed, a regression of the experimental data reveals that, in this region of the argon phase diagram, there is a quadratic dependency of the pressure on the density, highlighting that the isothermal compressibility is a rather sensitive function of the density itself. In summary, under moderate variations of the pressure, the LTh model is expected to be sufficiently accurate as density variations are expected to be rather small. Therefore, particular attention needs to be paid to density gradients emerging in non-equilibrium simulations targeting liquid conditions, in order to keep the LTh model within its range of applicability. In Fig. \ref{fig-PvsTc}, linear regressions are also calculated on the GenDPDE data, showing particularly that the estimated macroscopic isothermal compressibility, $\overline{\kappa}_T^*=1/1.02=0.98$, is satisfactorily close to the expected nominal value, $\overline{\kappa}_{T0}^*=1.00$, at the reference state point. Finally, considering the thermodynamic states in Fig. \ref{fig-PvsTc}-top, we evaluated the differences in internal energy per particle between adjacent points, with the aim of assessing the accuracy of the LTh model also in terms of internal energy. This analysis allows us to overcome the issue of considering different zero-energy levels than NIST, which prevented us from drawing a direct comparison with our numerical results for a single state point. Considering the points at $T_0^*$ and $T_0^*+\Delta T^*$, we find that the internal energy difference obtained from NIST data is $\Delta (U^*/N)^+ = -0.012$, against the $\Delta (U^*/N)^+ = -0.014$ from GenDPDE simulations. For the points at $T_0^*$ and $T_0^*-\Delta T^*$, experimental results yield $\Delta (U^*/N)^- = 0.013$, while from GenDPDE tests we obtain $\Delta (U^*/N)^- = 0.014$. These results confirm that the LTh model faithfully predicts the internal energy of the system.

\subsubsection{Analysis of supercritical conditions}

The above analysis shows that, when considering liquid conditions, the LTh model has a limited range of validity around the reference state point, due to the non-linear dependency of the system pressure with respect to the density. However, one would expect better agreement between numerical predictions and experimental data in regions where the fluid is more compressible, as the pressure is less sensitive to density variations. To check this, we considered a new reference state point for argon, characterized by a temperature $T_0 = 418.8$ K, mass density $\tilde{\rho}_0 = 695.99$ kg/m$^3$, and pressure $P_0 = 85.31$ MPa, which corresponds to supercritical conditions, well above the critical point \cite{NIST2024}. At these conditions, the specific macroscopic heat capacity is $\tilde{C}_{V0} = 3.56\times10^2$ J/(kg $\cdot$ K), the isothermal compressibility $\overline{\kappa}_{T0} = 6.83\times10^{-9}$ Pa\(^{-1}\), the thermal expansion coefficient $\overline{\alpha}_0 = 1.97\times10^{-3}$ K\(^{-1}\), the kinematic viscosity $\nu = 7.98 \times 10^{-8}$ m$^2$/s, and the thermal conductivity $\lambda = 0.0532$ W/(m $\cdot$ K). The corresponding dimensionless quantities and LTh parameters are reported in Table \ref{tab-refvals-SC}, obtained according to Sections \ref{Sub-LThpara} and \ref{Sub-CompDets}. Here, $\Rc^* = 2.1564$, as in the previous case.
\begin{table}[H]
    \centering
    \caption{Dimensionless state-point values and LTh parameters for GenDPDE simulations of supercritical argon at $T_0 = 418.8$ K, $\tilde{\rho}_0 = 695.99$ kg/m$^3$, and $P_0 = 85.31$ MPa.}
    %\vspace{2mm}
    %{\fontsize{7.6}{11}\selectfont
    \begin{tabularx}{\linewidth}{Y|Y|Y|Y|Y|Y|Y|Y|Y|Y|Y}
    \toprule
        $T_0^*$ & $P_0^*$ & $\occ_0^*$ & $C_{V0}^*$ & ${\kappa}_{T0}^*$ & ${\alpha}_{0}^*$ & $\theta_{0}^*$ & $\pi_{00}^*$ & $n_{00}^*$ & $\gamma^*$ & $\kappa^*$\\
    \midrule
        $0.0828$ & $0.583$ & $1.0$ & $8.56$ & $1.09$ & $9.78$ & $0.0828$ & $0.5211$ & $1.0$ & $1.05$ & $0.034$ \\
    \bottomrule
    \end{tabularx}
    %}
    \label{tab-refvals-SC}
\end{table}

Figure \ref{fig-Pvsc-SC} below shows the results of our GenDPDE simulations of supercritical argon, where the behavior of the LTh model is again analyzed in terms of measured pressure $P^*$, for imposed density variations $\Delta c_0^* = \pm 5\% \, \occ_0^*$ and $\Delta c_0^* = \pm 10\% \, \occ_0^*$. In comparison with the liquid case ({\em cf.} Fig. \ref{fig-PvsTc}), here we observe much better agreement with the experimental data from NIST, with errors below $3.5 \%$ in all cases (up to $2.5$ MPa).  %As before, the regression on the GenDPDE data shows that the calculated macroscopic isothermal compressibility, $\overline{\kappa}_T^*=1/1.06=0.94$, is still a good estimation of the nominal value, $\overline{\kappa}_{T0}^*=1.00$, at the reference state point.
These results confirm the wider range of applicability of the LTh model for fluids characterized by a more linear pressure-density relationship.
\begin{figure}[H]
    \centering
    \includegraphics[width=0.8\linewidth]{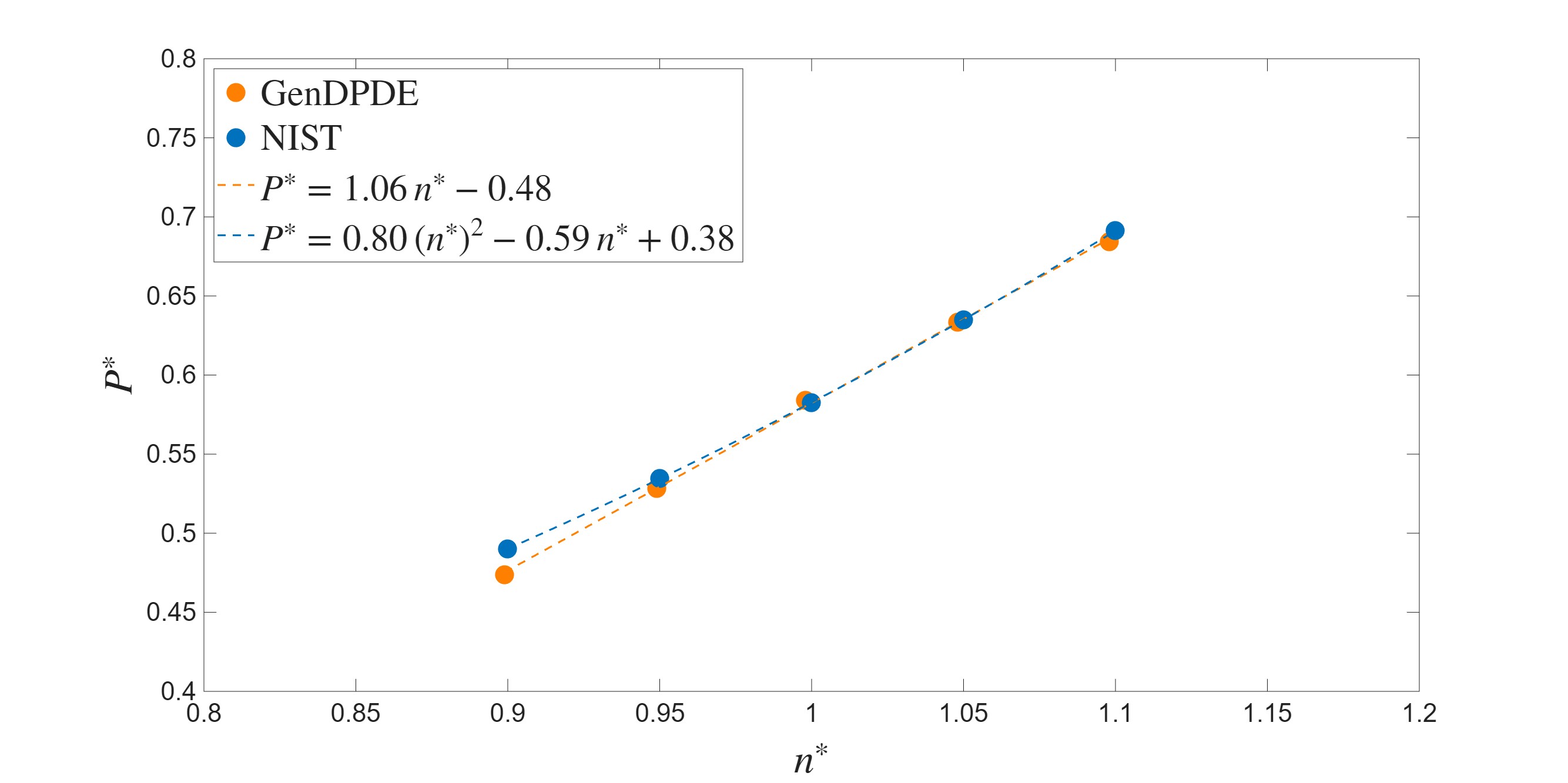}
    \caption{System pressure $P^*$ as a function of the number density $n^*$, as obtained from GenDPDE simulations of supercritical argon considering $\Rc^* = 2.1564$. The central point in the diagram refers to the reference state point. The numerical results are compared with data available from NIST \cite{NIST2024}.}
    \label{fig-Pvsc-SC}
\end{figure}

\section{Conclusions} \label{Sec-Fin}

In this work, we have exploited the flexibility of the Generalized Energy-Conserving Dissipative Particle Dynamics (GenDPDE) framework to develop a local thermodynamic (LTh) model for a general description of condensed phases at the mesoscale. Such a LTh model has been derived based on a set of a few relevant experimental parameters for the system, namely, the thermal expansion coefficient, $\alpha$, isothermal compressibility, $\kappa_T$, and constant-volume heat capacity, $C_V$, obtained at a given reference state. From these macroscopic parameters, we have shown a procedure to derive the mesoscopic counterparts, as the former are renormalized by the fluctuations in the particle state parameters ({\em cf.} Sec. \ref{Sub-LThpara}). 

In our analysis, we have considered argon as a paradigmatic test substance. Moreover, for simplicity, we have assumed that $\alpha$, $\kappa_T$ and $C_V$ are independent of particle properties such as temperature $\theta$ or density $n$. Despite this approximation, we have demonstrated the capability of the LTh model to reproduce the equilibrium thermodynamic properties of liquid argon, both at the reference state as well as for a significant range of temperatures and densities around it, with deviations up to $\pm 10\%$ for the former and $\pm 5\%$ for the latter, which permits the analysis of non-equilibrium scenarios involving sufficiently wide temperature and concentration differences. We have furthermore extended the investigation of the model's range of validity to the analysis of supercritical conditions, which are of interest in many engineering applications. We have shown that, under such conditions, the LTh model is able to accurately estimate the targeted thermodynamic states for an even wider range of density deviations as compared to liquid phase, namely, up to $\pm 10\%$, given that $\kappa_T$ is independent of the density over wider regions in supercritical fluids. 

From a theoretical perspective, we have also derived approximate analytical predictions for the macroscopic Equations of State (EoS) from the mesoscopic parameters associated to the LTh model, whose determination is a cornerstone of the proposed LTh model. Hence, these analytical expressions establish the bridge between the experimentally accessible macroscopic coefficients and the adequate mesoscopic parameters to be used in the simulations. To this end, we have proposed an expansion of the many-body interparticle potential in density fluctuations \cite{Merabia2007}. Our analysis reveals that, even at the lowest level of approximation, local particle arrangements play a relevant role in determining the macroscopic properties of the system, through the equilibrium distribution functions $g$. More specifically, we have found that unavoidably both two-body ($g(\br'|\br)$) and three-body ($g(\br''|\br, \br')$) correlations need to be accounted for, when estimating the system pressure $P$ from the mesoscopic model, to eventually compare with the simulation results. For systems in which three-body correlations are subdominant, the Kirkwood Superposition Approximation \cite{Kirkwood1952} provides a valid route to describe these interactions in terms of pairwise contributions, which we obtain from the simulations. Nevertheless, when correlations become stronger, the approximations underlying our EoS predictions start to fall outside their range of applicability. We have demonstrated the validity and limitations of the theoretical estimations by considering different values of the cutoff radius $\Rc$ for the many-body temperature-dependent potential interactions in our simulations of liquid argon. In particular, we have found that, at small $\Rc$, the radial distribution function $g(r)$ obtained from GenDPDE tests is characterized by strong short-range correlations. In this case, the parametrization using the derived EoS produces less accurate estimates of the system pressure and internal energy. For larger $\Rc$, on the other hand, the accuracy of the analytical predictions is excellent, as the local density fluctuations decrease as $\Rc^{-3}$.

For completeness, we have furthermore investigated the applicability of the Hypernetted Chain (HNC) approximation as a tool to predict the local structure of our mesoscopic system embedded in a theoretical $g(r)$. In all the considered cases, we have found very good agreement between the $g(r)$ obtained from GenDPDE tests and the HNC predictions. Thus, the expansion of the density- and temperature-dependent potential, together with the Kirkwood approximation and the HNC strategy, constitute a powerful set of tools for the analysis of the structure of systems responding to this type of many-body potentials. Nevertheless, when the HNC results are used as input in the macroscopic EoS, the resulting estimates of the thermodynamic quantities are systematically less accurate than those obtained using the $g(r)$ from GenDPDE simulations, with discrepancies that increase as $\Rc$ decreases. This indicates that, in the present framework, HNC is more effective as a tool for structural analysis than as a quantitative tool for parameter determination. For accurate liquid-state simulations, the final calibration of the LTh parameters is thus most reliably performed using a simulation-informed refinement procedure.

The LTh model developed herein provides a reliable basis for the analysis of liquids and supercritical fluids at the mesoscopic level, both qualitatively and quantitatively, using CG techniques. Due to the applicability of GenDPDE to the analysis of non-equilibrium situations, phenomena such as, e.g., thermophoresis could thus be addressed, considering both simple systems as well as more complex configurations, including colloidal suspensions and polymeric structures, with important implications in academic and industrial contexts. 

\begin{acknowledgement}
GC acknowledges funding from the European Union's Horizon 2020 research and innovation programme under the Marie Skłodowska-Curie grant agreement No. 945413 (DOI 10.3030/945413) and from the Universitat Rovira i Virgili (URV). AM and JBA acknowledge support from the grant PID2021-122187NB-C33, funded by MCIN\slash AEI\slash 10.13039\slash 501100011033 and “ERDF A way of making Europe”, and the grant PID2024-158902NB-I00, funded by MICIU\slash AEI\slash 10.13039\slash 501100011033 and by “ERDF A way of making Europe”. Research performed by AM and JBA was furthermore sponsored by the Army Research Office (DOI 10.13039/100000183), and was accomplished under Cooperative Agreement No. W911NF-20-2-0227. FB acknowledges funding from the Royal Society through the International Exchanges Programme (Grant number IES\textbackslash R3\textbackslash 193119).
\end{acknowledgement}

\begin{disclaimer}
This work reflects only the authors’ view and the Agency is not responsible for any use that may be made of the information it contains. The views and conclusions contained in this document are those of the authors and should not be interpreted as representing the official policies, either expressed or implied, of the Army Research Office or the U.S. Government. The U.S. Government is authorized to reproduce and distribute reprints for Government purposes notwithstanding any copyright notation herein.
\end{disclaimer}

%\clearpage
\appendix

\section{Factorization of the partition function}\label{App-QN}

Given the particular form of the particle EoS, Eqs. \eqref{us} and \eqref{genpi}, the integration over the particle entropy in the partition function Eq. \eqref{partition2} can be analytically evaluated. Effectively, by defining the new variable,
\begin{align}
 z \equiv \frac{C_V \Theta}{k_BT} \left[e^{s/k_B} \psi	\right]^{\frac{k_B}{C_V}} \label{zvar}
 %; \;\;\;  ds = C_V \, \frac{dz}{z}
\end{align}
the integration over entropy in Eq. \eqref{partition2} reads
\begin{align}
&\int_0^\infty ds \, \Theta \,e^{\left(Ts-C_V \Theta \,  e^{s_/C_V}[\psi(n)]^{\frac{k_B}{C_V}}- {\cal V}(n) \right)/k_BT} = \nonumber\\
&= \frac{C_V \Theta}{ \psi(n)} \left(\frac{k_BT}{C_V \Theta}	\right)^{\frac{C_V}{k_B}} \, e^{-\frac{{\cal V}(n)}{k_BT}} \int_0^\infty dz \, z^{\left( \frac{C_V}{k_B}-1 \right)} \,  e^{-z} = \nonumber \\
& = \frac{C_V \Theta}{ \psi(n)} \left(\frac{k_BT}{C_V \Theta}	\right)^{\frac{C_V}{k_B}} \, e^{-\frac{{\cal V}(n)}{k_BT}}  \, \Gamma \left( \frac{C_V}{k_B} \right)		\label{sint}
\end{align}
Hence, the partition function can be written as
\begin{align}
Q_N(V,T) =& \,\ \frac{(C_V \Theta)^N}{\Lambda^{3N} \varepsilon^N \,N!}  \left(\frac{k_BT}{C_V \Theta}	\right)^{\frac{NC_V}{k_B}} \, \Gamma^N \left(\frac{C_V}{k_B} \right) \times \nonumber \\
& \times \int d\br_1 \dots d\br_N \, e^{-\sum_i \left[\frac{\mathcal{V}(n_i)}{k_BT}+\ln \psi(n_i) \right]} \label{partitionapp4}
\end{align}
The remaining integral is a so-called configurational integral, where the potential contains an additional temperature-independent contribution.

At this point, it is useful to introduce an ideal-gas limit for the mesoparticles, obtained by setting $\mathcal{V}(n_i)/{k_BT}+\ln \psi(n_i)=0$. For this case, the partition function reads,
\begin{align}
Q_N^{\text{gas}}(V,T) = &\frac{(C_V \Theta)^N}{\Lambda^{3N} \varepsilon^N \,N!}  \left(\frac{k_BT}{C_V \Theta}	\right)^{\frac{NC_V}{k_B}} \, \Gamma^N \left(\frac{C_V}{k_B} \right) V^N 
	\label{partitionapp3}
\end{align}
The contribution due to the internal energy of the particles can be further separated from the classical ideal-gas contribution, yielding,
\begin{align}
Q_N^{\text{gas}}(V,T) &= \left[	\frac{C_V}{k_B} 	\left(\frac{k_BT}{C_V \Theta} \right)^{\frac{C_V}{k_B}} \Gamma \left(\frac{C_V}{k_B} \right)\right]^N \left(\frac{V \, e}	{\Lambda^3 N}	\right)^N \nonumber\\
&\equiv [q(T)]^N \, [Q^{id}(T,V)]^N \label{partitionapp5}
\end{align}
where Stirling's approximation has been used, along with Eq. \eqref{Debye}. Here, $Q^{id}(T,V)$ stands for the ideal-gas partition function as usually defined, i.e.,
\begin{align}
Q^{id}(V,T) \equiv \frac{V \, e}{\Lambda^3 N} \label{idealgas}
\end{align}
whereas the internal energy contribution reads,
\begin{align}
q(T) \equiv \frac{C_V}{k_B} 	\left(\frac{k_BT}{C_V \Theta}	\right)^{\frac{C_V}{k_B}} \,\Gamma \left(\frac{C_V}{k_B} \right)	\label{qapp}
\end{align}
This permits us to write the partition function as the product of three terms, namely,
\begin{align}
Q_N(V,T) = [q(T)]^N \, [Q^{id}(V,T)]^N \, Q_N^{ex}(V,T) \label{factorapp}
\end{align}
with the excess partition function given by,
\begin{equation}
Q_N^{ex}(V,T) \equiv \frac{1}{V^N} \int d\br_1 \dots d\br_N \, e^{-\sum_i \left[\frac{\mathcal{V}(n_i)}{k_BT}+\ln \psi(n_i) \right]}  \label{partitionapp6}
\end{equation}

\section{The excess pressure} \label{App-Merabia}

In the canonical ensemble, the excess pressure is related to the excess partition function, Eq. \eqref{partition6}, as,
\begin{align}
    P^\text{ex} = \left. -\frac{\partial }{\partial V} F^\text{ex} \right|_T =\left. k_BT  \frac{\partial }{\partial V} \ln Q_N^\text{ex} \right|_T
\end{align}
Following Ref. \cite{ChandlerBook}, we rescale the coordinates to $\bx_i \equiv \br_i V^{-1/3}$, so that the limits of integration become independent of volume variations. Thus,
\begin{align}
    P^\text{ex} = & k_BT \frac{\partial }{\partial V} \ln \int d\bx_1\dots d\bx_N \, e^{-\sum_i \mathcal{W}(T,n_i)/k_BT} = \nonumber \\
    = &k_BT \frac{1}{Q_{N}^\text{ex}} \int d\bx_1\dots d\bx_N \, \frac{\partial }{\partial V}e^{-\sum_i \mathcal{W}(T,n_i)/k_BT}
\end{align}
Differentiating the integrand with respect to the volume, and taking into account that, for a given particle conformation, 
\begin{align}
    \frac{\partial n^b_i}{\partial V} = & \sum_{j \neq i} w_{ij}' \, \frac{\partial r_{ij}}{\partial V} = \sum_{j \neq i} w_{ij}' \, x_{ij} \frac{1}{3} V^{-2/3} = \nonumber \\
    = & \frac{1}{3V} \sum_{j \neq i} w_{ij}' \, r_{ij}
\end{align}
we readily arrive at 
\begin{align}
     P^\text{ex} = & -\frac{1}{3V}\left\langle \sum_{i, j\neq i} \mathcal{W}_n(T,n_i) \zeta_i w_{ij}' \, r_{ij}\right \rangle = \nonumber \\
     = & -\frac{1}{3V}\left\langle \sum_{i, j<i} \bigg(\frac{\pi_i}{n_i^2} \zeta_i + \frac{\pi_j}{n_j^2} \zeta_j \bigg) w_{ij}' \, r_{ij}\right \rangle \label{virialAPP}
\end{align}
where use of Eq. \eqref{difw} has been made to obtain the last equality. Using the notation introduced in Eq. \eqref{Wn}, Eq. \eqref{virial} can be reformulated as ({\em cf.} Eq. \eqref{Pex})
\begin{align}
    P^{\text{ex}} =& \, -  \frac{1}{6V} \left\langle \sum_{i, j \neq i} \bigg([\mathcal{W}_n]_{\hat{n}_i^b} + [\mathcal{W}_n]_{\hat{n}_j^b} \bigg) r_{ij} w'(r_{ij}) \right\rangle \label{PexAPP}
\end{align}
For the sake of clarity and without loss of generality, we now focus only on the first of the two contributions in Eq. \eqref{PexAPP}, namely,
\begin{align}
    P^{\text{ex}}_{(1)} =& \, -  \frac{1}{6V} \left\langle \sum_{i, j \neq i} [\mathcal{W}_n]_{\hat{n}_i^b} r_{ij} w'(r_{ij}) \right\rangle \label{PexAPP2}
\end{align}
Using the expansion of $\mathcal{W}_n$ Eq. \eqref{Wnexp}, Eq. \eqref{PexAPP2} reads,
%
%\begin{widetext}
\begin{align}
    P^{\text{ex}}_{(1)} \simeq & \, -  \frac{1}{6V} \bigg\langle \sum_{i, j \neq i} \bigg( [\mathcal{W}_n]_{n(\br_i)} - [\mathcal{W}_{nn}]_{n^b(\br_i)} n^b(\br_i) \bigg) r_{ij} w'(r_{ij}) \,\ + \nonumber \\ 
    & \phantom{-  \frac{1}{6V} \bigg\langle \,\ } + \sum_{i, j \neq i, k \neq i} [\mathcal{W}_{nn}]_{n^b(\br_i)} \, r_{ij} w(r_{ik}) w'(r_{ij}) \bigg\rangle \label{PexAPP3}
\end{align}
%\end{widetext}
%
In the second summation in this expression, the terms considering $k=j$ can be conveniently singled out, yielding,
\begin{align}
    \sum_{i, j \neq i, k \neq i} & [\mathcal{W}_{nn}]_{n^b(\br_i)} \, r_{ij} w(r_{ik}) w'(r_{ij}) = \nonumber \\
    =& \sum_{i, j \neq i, k \neq j} [\mathcal{W}_{nn}]_{n^b(\br_i)} \, r_{ij} w(r_{ik}) w'(r_{ij}) \,\ + \nonumber \\
    &+ \sum_{i, j \neq i}[\mathcal{W}_{nn}]_{n^b(\br_i)} \, r_{ij} w(r_{ij}) w'(r_{ij}) \label{sumijk}
\end{align}
Thus, introducing the instantaneous densities $\hat{c}(\br) \equiv \sum_i \delta(\br-\br_i)$, $\hat{c}^*(\br) \equiv \sum_{j \neq i} \delta(\br-\br_j)$, and $\hat{c}^{**}(\br) \equiv \sum_{k \neq i,j} \delta(\br-\br_k)$, Eq. \eqref{PexAPP3} can be written as,
%
%\begin{widetext}
\begin{align}
    P^{\text{ex}}_{(1)} \simeq & - \frac{1}{6V} \int d\br d\br' \bigg( [\mathcal{W}_n]_{n(\br)} - [\mathcal{W}_{nn}]_{n^b(\br)} n^b(\br) \bigg)|\br-\br'| w'(|\br-\br'|) \big\langle\hat{c}(\br)\hat{c}^*(\br')\big\rangle \,\ + \nonumber \\
    & -  \frac{1}{6V} \int d\br d\br' [\mathcal{W}_{nn}]_{n^b(\br)} \, |\br-\br'| w(|\br-\br'|) w'(|\br-\br'|) \big\langle\hat{c}(\br)\hat{c}^*(\br')\big\rangle \,\ + \nonumber \\
    & - \frac{1}{6V} \int d\br d\br' d\br'' [\mathcal{W}_{nn}]_{n^b(\br)} \, |\br-\br'| w(|\br-\br''|) w'(|\br-\br'|) \big\langle\hat{c}(\br)\hat{c}^*(\br')\hat{c}^{**}(\br'')\big\rangle = \nonumber \\
    =& - \frac{1}{6V} \int d\br d\br' \bigg( [\mathcal{W}_n]_{n(\br)} - [\mathcal{W}_{nn}]_{n^b(\br)} n^b(\br) \bigg)|\br-\br'| w'(|\br-\br'|) c(\br)c(\br')g(\br'|\br) \,\ + \nonumber \\
    & -  \frac{1}{6V} \int d\br d\br' [\mathcal{W}_{nn}]_{n^b(\br)} \, |\br-\br'| w(|\br-\br'|) w'(|\br-\br'|) c(\br)c(\br')g(\br'|\br) \,\ + \nonumber \\
    & - \frac{1}{6V} \int d\br d\br' d\br'' [\mathcal{W}_{nn}]_{n^b(\br)} \, |\br-\br'| w(|\br-\br''|) w'(|\br-\br'|) c(\br)c(\br')c(\br'')g(\br''|\br,\br') \label{PexfieldsAPP}
\end{align}
where $g(\br''|\br,\br')$ is the triplet distribution function. Exploiting the definition of $n^b(\br)$ Eq. \eqref{local}, this last expression can be further manipulated to obtain
\begin{align}
    P^{\text{ex}}_{(1)} =& - \frac{1}{6V} \int d\br d\br' [\mathcal{W}_n]_{n(\br)} |\br-\br'| w'(|\br-\br'|) c(\br)c(\br')g(\br'|\br) \,\ + \nonumber \\
    & - \frac{1}{6V} \int d\br d\br'[\mathcal{W}_{nn}]_{n^b(\br)} |\br-\br'| w(|\br-\br'|) w'(|\br-\br'|) c(\br)c(\br')g(\br'|\br) \,\ + \nonumber \\
    &-  \frac{1}{6V} \int d\br d\br' d\br'' [\mathcal{W}_{nn}]_{n^b(\br)} |\br-\br'| w(|\br-\br''|) w'(|\br-\br'|) c(\br)c(\br')c(\br'') \times \nonumber \\
    & \phantom{-  \frac{1}{6V} \int d\br d\br' d\br'' \,\ } \times \left[g(\br''|\br,\br')-g(\br'|\br)g(\br''|\br) \right] \label{PexfieldsAPP2}
\end{align}
Carrying out an analogous derivation for the second term in Eq. \eqref{PexAPP}, one gets to the same result as in Eq. \eqref{PexfieldsAPP2}. Thus, the total excess pressure is simply twice $P^{\text{ex}}_{(1)}$, and reads ({\em cf.} Eq. \eqref{Pexfields}), 
\begin{align}
    P^{\text{ex}} =& - \frac{1}{3V} \int d\br d\br' [\mathcal{W}_n]_{n(\br)} |\br-\br'| w'(|\br-\br'|) c(\br)c(\br')g(\br'|\br) \,\ + \nonumber \\
    & - \frac{1}{3V} \int d\br d\br'[\mathcal{W}_{nn}]_{n^b(\br)} |\br-\br'| w(|\br-\br'|) w'(|\br-\br'|) c(\br)c(\br')g(\br'|\br) \,\ + \nonumber \\
    &-  \frac{1}{3V} \int d\br d\br' d\br'' [\mathcal{W}_{nn}]_{n^b(\br)} |\br-\br'| w(|\br-\br''|) w'(|\br-\br'|) c(\br)c(\br')c(\br'')\times \nonumber \\
    & \phantom{-  \frac{1}{3V} \int d\br d\br' d\br'' \,\ } \times \left[g(\br''|\br,\br')-g(\br'|\br)g(\br''|\br) \right] \label{PexfieldsAPP3}
\end{align}
%\end{widetext}
%

\section{Potential contribution to the internal energy} \label{App-U}

To evaluate the potential contribution to the internal energy of the system, we start from Eq. \eqref{avgu},

\begin{align}
     \Big\langle \sum_i u(s_i,\hat{n}_i)\Big\rangle = \sum_i \int & d\br^N ds^N P_{\text{eq}}(\br^N,s^N) \Big[  \mathcal{V}(\hat{n}_i) \,\ + \nonumber \\ &+ C_V\Theta e^{s_i/C_V}[\psi(\hat{n}_i)]^{k_B/C_V} \Big] \label{avguAPP}
\end{align}
Let us initially focus on the second term in this integral, namely,
%
%\begin{widetext}
\begin{align}
    \sum_i\int & d\br^N ds^N P_{\text{eq}}(\br^N,s^N) C_V\Theta e^{\frac{s_i}{C_V}}[\psi(\hat{n}_i)]^{\frac{k_B}{C_V}} = \nonumber \\
    &= \sum_i \int d\br^N ds^N \frac{1}{q^N \tilde{Q}^{ex}_N} e^{-\sum_j\Big[ \mathcal{V} (\hat{n}_j) +  C_V\Theta e^{\frac{s_j}{C_V}}[\psi(\hat{n}_j)]^{\frac{k_B}{C_V}}-Ts_j\Big]/k_BT} C_V\Theta e^{\frac{s_i}{C_V}}[\psi(\hat{n}_i)]^{\frac{k_B}{C_V}} \label{psitermAPP}
\end{align}
%\end{widetext}
%
where $q$ and $\tilde{Q}^{ex}\equiv V^NQ^{ex}$ are the internal and excess partition functions, Eqs. \eqref{qapp} and \eqref{partitionapp6} respectively. Here, the momentum contribution has already been integrated out for simplicity. Proposing a change of variable analogous to Eq. \eqref{zvar}, Eq. \eqref{psitermAPP} becomes,
%
%\begin{widetext}
\begin{align}
    \sum_i & \int d\br^N ds^N \frac{1}{q^N \tilde{Q}^{ex}_N} e^{-\sum_j\Big[ \mathcal{V} (\hat{n}_j) +  C_V\Theta e^{\frac{s_j}{C_V}}[\psi(\hat{n}_j)]^{\frac{k_B}{C_V}}-Ts_j\Big]/k_BT} C_V\Theta e^{\frac{s_i}{C_V}}[\psi(\hat{n}_i)]^{\frac{k_B}{C_V}} = \nonumber \\
    &= \sum_i \int d\br^N dz^N \frac{C_V^N}{\Pi_k \, z_k} \frac{1}{q^N\tilde{Q}^{ex}_N} k_BT \, z_i \, \,\ \times \nonumber \\
    & \phantom{= \sum_i \int d\br^N dz^N} \times e^{-\sum_j \Big[ \mathcal{V} (\hat{n}_j) /k_BT \, + \, z_j \, - \,  (C_V/k_B)\ln(z_j)\, + \, \ln(\psi(\hat{n}_j)) \, - \, (C_V/k_B)\ln[(k_BT)/(C_V\Theta)] \Big]} = \nonumber \\
    &= \sum_i \int d\br^N \frac{1}{\tilde{Q}_N^{ex}} e^{\sum_j \Big[ \mathcal{V} (\hat{n}_j) /k_BT \, + \, \ln(\psi(\hat{n}_j)) \Big]} \,\ \bigg(\frac{k_BT}{C_V\Theta}\bigg)^\frac{Nk_B}{C_V} \,\ \times \nonumber \\
    & \phantom{= \sum_i \int d\br^N dz^N} \times C_V^N k_BT  \int dz^N \frac{1}{q^N} \frac{z_i}{\Pi_k \, z_k} e^{-\sum_j \Big[z_j \, - \,  (C_V/k_B)\ln(z_j)\Big]} \label{psitermAPP2}
\end{align}
As the positional integral in Eq. \eqref{psitermAPP2} is equal to $1$, we are thus left with the evaluation of the entropic part,
\begin{align}
    k_BT \, & C_V^N \bigg(\frac{k_BT}{C_V\Theta}\bigg)^\frac{Nk_B}{C_V} \sum_i \int_0^\infty dz^N \frac{1}{q^N} \frac{z_i}{\Pi_k \, z_k} e^{-\sum_j \Big[z_j \, - \,  (C_V/k_B)\ln(z_j)\Big]} = \nonumber \\
    &= k_BT \, C_V\bigg(\frac{k_BT}{C_V\Theta}\bigg)^\frac{k_B}{C_V} \frac{q^{N-1}}{q^N} \sum_i \Gamma\Big(\frac{C_V}{k_B}+1 \Big) = \nonumber \\
    &= Nk_BT \frac{\Gamma\Big(\frac{C_V}{k_B}+1 \Big)}{\Gamma\Big(\frac{C_V}{k_B} \Big)} = \nonumber \\
    &= NC_VT \label{psitermAPP3}
\end{align}
which corresponds to the result in Eq. \eqref{psiterm}. Next, we focus on the first contribution in Eq. \eqref{avguAPP},
\begin{align}
    \sum_i \int & d\br^N ds^N P_{\text{eq}}(\br^N,s^N) \mathcal{V}(\hat{n}_i) \simeq \nonumber \\
    &\simeq \sum_i \int d\br^N P_{\text{eq}}(\br^N) \Big[ \mathcal{V}(n(\br_i)) - [\mathcal{V}_n]_{n^b(\br_i)} n^b(\br_i) + [\mathcal{V}_n]_{n^b(\br_i)} \sum_{j \neq i} w(|\br_i - \br_j|) \Big] \label{calVtermAPP}
\end{align}
where use of the expansion Eq. \eqref{calVexpansion} has been made. For a homogeneous system, Eq. \eqref{calVtermAPP} can be reformulated as,
\begin{align}
    \sum_i \int & d\br^N P_{\text{eq}}(\br^N) \Big[ \mathcal{V}(n) - [\mathcal{V}_n]_{n^b} n^b + [\mathcal{V}_n]_{n^b} \sum_{j \neq i} w(|\br_i - \br_j|) \Big] = \nonumber \\
    &= \sum_i \bigg[ \mathcal{V}(n) - [\mathcal{V}_n]_{n^b} n^b + [\mathcal{V}_n]_{n^b} \int d\br^N P_{\text{eq}}(\br^N)\sum_{j \neq i} w(|\br_i - \br_j|)\bigg] = \nonumber \\
    &= N\mathcal{V}(n) - N[\mathcal{V}_n]_{n^b} n^b + [\mathcal{V}_n]_{n^b}\int d\br^N P_{\text{eq}}(\br^N)\sum_i\sum_{j \neq i} w(|\br_i - \br_j|) = \nonumber \\
    &= N\mathcal{V}(n) - N[\mathcal{V}_n]_{n^b} n^b + [\mathcal{V}_n]_{n^b}\int d\br d\br' w(|\br - \br'|) \Big\langle \sum_i\sum_{j \neq i} \delta(\br-\br_i)\delta(\br'-\br_j) \Big\rangle = \nonumber \\
    &= N\mathcal{V}(n) - N[\mathcal{V}_n]_{n^b} n^b + N [\mathcal{V}_n]_{n^b} \, \occ\int d\Delta\br \,  w(\Delta r) g(\Delta \br) \label{calVtermAPP2}
\end{align}
%\end{widetext}
%
where we have introduced the instantaneous fields to obtain the last equality. The third contribution in the last line of Eq. \eqref{calVtermAPP2} is equal to $N[\mathcal{V}_n]_{n^b}n^b$ ({\em cf.} Eq. \eqref{local2}), so that it identically cancels the second contribution and we finally obtain,
\begin{align}
    \sum_i \int & d\br^N ds^N P_{\text{eq}}(\br^N,s^N) \mathcal{V}(\hat{n}_i) = N\mathcal{V}(n) \label{calVtermAPP3}
\end{align}
as reported in Eq. \eqref{calVterm}.

\bibliography{references}

\end{document}